\documentclass[a4paper,12pt]{article}

\title{Superintegrable Systems in Darboux spaces}

\author{E.~G.~Kalnins\thanks{Email: e.kalnins@waikato.ac.nz} \\
{\small Department of Mathematics, University of Waikato,} \\[-4pt]
{\small Private Bag 3105, Hamilton, New Zealand.}
\\ \\
J.~M.~Kress\thanks{Email: jonathan@maths.unsw.edu.au} \\
{\small School of Mathematics, The University of New South Wales,} \\[-4pt]
{\small Sydney NSW 2052, Australia} \\[-4pt]
\\ \\
W.~Miller,~Jr.\thanks{Email: miller@ima.umn.edu} \\
{\small School of Mathematics, University of Minnesota,} \\[-4pt]
{\small Minneapolis, Minnesota, 55455, U.S.A.} \\[-4pt]
\\ \\
P.~Winternitz\thanks{Email: wintern@crm.umontreal.ca} \\
{\small Centre de Recherches Math\'ematiques et D\'epartement de Math\'ematiques   
et de Statistique} \\[-4pt]
{\small Universit\'e de Montr\'eal,} \\[-4pt]
{\small C.P. 6128-CV, Montr\'eal, Qu\'ebec H3C 3J7, Canada.}}

\usepackage{amssymb}

\newcounter{commentnumber}
\usepackage{calc}
\usepackage{ifthen}
\newcounter{hour}\newcounter{minute}
\ifthenelse{\theminute<10}
   {}
   {}

\newcommand{\bibtitle}[1]{#1}

\newcommand{\be}{\begin{equation}}
\newcommand{\ee}{\end{equation}}
\newcommand{\ba}{\begin{array}}
\newcommand{\ea}{\end{array}}
\newcommand{\bea}{\begin{eqnarray}}
\newcommand{\eea}{\end{eqnarray}}
\newcommand{\beann}{\begin{eqnarray*}}
\newcommand{\eeann}{\end{eqnarray*}}
\newcommand{\bi}{\begin{itemize}}
\newcommand{\ei}{\end{itemize}}

\newcommand{\ds}{\displaystyle}

\newcommand{\sech}{\mbox{sech}}
\newcommand{\cosec}{\mbox{cosec}}
\newcommand{\cosech}{\mbox{cosech}}
\newcommand{\pd}[2]{\frac{\partial{#1}}{\partial{#2}}}
\newcommand{\pdtwo}[2]{\frac{\partial^2{#1}}{\partial{#2}^2}}
\newcommand{\pdsq}[2]{\left(\ds\pd{#1}{#2}\right)^2}
\newcommand{\sfrac}[2]{\mbox{$\frac{#1}{#2}$}}

\date{\today}

\makeatletter
\@addtoreset{equation}{section}
\renewcommand{\theequation}{\thesection.\@arabic\c@equation}
\makeatother

\begin{document}

\maketitle

\begin{abstract}
Almost all research on superintegrable potentials concerns spaces of
constant curvature. In this paper we find by
exhaustive calculation, all superintegrable potentials in the four
Darboux spaces of revolution that have at least two integrals of
motion quadratic in the momenta, in addition to the Hamiltonian. These
are two-dimensional spaces of nonconstant curvature. It turns out that
all of these potentials are equivalent to superintegrable potentials
in complex Euclidean 2-space or on the complex 2-sphere, via
``coupling constant metamorphosis'' (or equivalently, via
St\"ackel multiplier transformations). We present tables of the results.
\end{abstract}

\newpage

\section{Introduction}
\label{sec:intro}

In a previous article \cite{KKW} we have studied superintegrability 
in a two-dimensional 
space of nonconstant curvature, in particular one of the so 
called Darboux spaces, given by Koenigs \cite{Koenigs}. 
In this article we study the remaining 
three spaces of nonconstant curvature from the point of view of 
superintegrability. This involves the addition of a potential to each of the 
spaces given by Koenigs. We recall that classical superintegrability relating to
a Hamiltonian $H(x_1,\ldots,x_n,p_1,\ldots,p_n)=H(x,p)$ implies the existence
of $2n-1$  
globally defined constants of the motion. For the purposes of this article we 
restrict this definition to require that there exist $2n-1$ globally
defined functionally independent
constants of the motion $X_i$, $i=1,\ldots,2n-1$  that are quadratic
in the canonical  momenta $p_i$. This clearly implies the relations  
$$
\{H,X_\ell\} = \sum^n_{i=1}
   \left(\pd{X_\ell}{x_i}\pd H{p_i}-\pd{X_\ell}{p_i}\pd H{x_i}\right)
 = 0\,, \qquad i=1,\ldots,2n-1\,.
$$

The concepts of integrability and superintegrability also have their analogue 
in quantum mechanics. A superintegrable quantum mechanical system is 
described by $2n-1$ (independent) quantum observables 
$\hat H=\hat X_1,\hat X_2,\ldots,\hat X_{2n-1}$ that satisfy the commutation 
relations  
$$
[\hat H,\hat X_i]= \hat H\hat X_i-\hat X_i\hat H=0 ,\qquad
i=1,\ldots,2n-1.
$$
The analogue of quadratic superintegrability in this case  is
that each of the quantum observables is a second order partial differential 
operator. Systematic studies of superintegrable systems have been conducted in 
spaces of constant curvature in two  dimensions 
\cite{FMSUW, WSUF, KMJP2, KMJP3, KKMP}.

In this article we solve the following problem.
Given a Riemannian space in two dimensions with infinitesimal distance  
$
ds^2=\sum_{i,j=1}^2g_{ij}(u)du^idu^j
$,
and $u =(u^1,u^2)$, the classical Hamiltonian has the form  
$$
H=\sum_{i,j=1}^2g^{ij}p_ip_j+V(u)
$$
and the corresponding Schr\"odinger equation is
$$
\hat H\Psi =\frac{1}{\sqrt{g}}\partial _{u^i}\left(\sqrt{g }
g^{ik}\partial _{u^k}\Psi\right)+V(u)\Psi =E\Psi 
$$
where $\sqrt{ g} = \det (g_{ij})$. Koenigs  found  all free Hamiltonians  
$H=\sum g^{ij}p_ip_j$
admitting at least two extra functionally independent constants of the motion
of the form 
$$
\Lambda  = \sum_{i,j=1}^2a^{ij}(u)p_ip_j,\qquad a^{ij}= a^{ji}.
$$
He obtained a number of families of solutions; in particular,  spaces
that admitted three extra quadratic constants. There must then be a
functional relation between these and, furthermore, in each case there is a 
Killing vector, i.e., a function $\mu =\sum_{i=1}^2a^i(u)p_i$ that satisfies  
$\{H,\mu \}=0$.  One of the three quadratic constants is a square of the Killing
vector $\mu$.

The problem we solve here is supplemental to that of Koenigs:  Suppose we have a 
Hamiltonian  $H=\sum g^{ij}p_ip_j+V(u)$
that admits a Killing vector.  We determine the {\it potentials} that correspond to
superintegrability, i.e., potentials such that we can find at least two 
extra functionally independent quadratic constants of the form 
$$
\Lambda  = \sum_{i,j=1}^2a^{ij}(u)p_ip_j+\lambda (u)\,.
$$
A necessary condition that this be possible is that the Riemannian
space be one of the four listed by Koenigs:
\begin{enumerate}
\item  $\ds ds^2=(x+y)dxdy$
\item  $\ds ds^2=\left({a\over (x-y)^2} +b\right)dxdy$
\item  $\ds ds^2=\left(ae^{-(x+y)/2}+be^{-x-y}\right)dxdy$
\item  $\ds ds^2= \frac{a\left(e^{(x-y)/2} +e^{(y-x)/2}\right)+b}
                       {\left(e^{(x-y)/2}-e^{(y-x)/2}\right)^2}\, dxdy$.
\end{enumerate}

The first of these spaces, type one, or $D_1$, has been treated in detail 
in an earlier paper \cite{KKW}.  Here we treat the remaining three Darboux spaces
in a similar and unified way.  Sections \ref{sec:Darboux2}, 
\ref{sec:Darboux3}, and \ref{sec:Darboux4} are devoted to the 
spaces $D_2$, $D_3$, and $D_4$, respectively. In each space we follow the same 
pattern.

\begin{enumerate}
\item
We first consider a classical free particle system and give the free
Hamiltonian $H_0$, the Killing vector $K$ and the two Killing tensors $X_1$ and
$X_2$ in a space with a conformally Euclidean metric (real or complex). We
choose coordinates $u$ and $v$ in which the first order constant is $K = p_v$,
 hence $u$ is an ignorable variable, not appearing in the metric or in
the Hamiltonian.
\item
We present an embedding of the two-dimensional Darboux space into a 
three-dimensional flat space.
\item
We present a polynomial relation between the four integrals of motion
$H$, $K$, $X_1$ and $X_2$, and also the polynomial algebra generated by these 
integrals.
\item
We consider the quantum mechanics of a free particle in the 
corresponding Darboux space, i.e., write the corresponding Hamiltonian and 
integrals of motion as linear operators. We then establish that the 
relations between these operators are the same as those between the 
classical quantities.
\item
We use the fact that the Killing vector $K$ generates a one-dimensional
Lie transformation group to classify all integrals of motion
\be
\lambda = a X_1 + b X_2 + c K^2
\ee
into conjugation classes. Each class gives rise to a coordinate system in
which the Hamilton-Jacobi and Schr\"odinger equations allow the
separation of variables. We construct these separable coordinate systems
explicitly and solve the corresponding separated equations (classical and
quantum).
\item
By construction, the free classical and quantum systems in Darboux
spaces are all quadratically superintegrable: they have 3 functionally
independent integrals of motion. We introduce potentials that do not 
destroy this superintegrability. Thus we present systematically  all 
superintegrable classical and quantum systems of the form
\be
H = H_0  + V(u,v)
\ee
where $H_0$ is the free Hamiltonian in the space $D_2$, $D_3$, or $D_4$. To obtain 
this result we make use of the fact that to be quadratically 
superintegrable, a Hamiltonian in a Darboux space must allow the 
separation of variables in at least two coordinate systems.
\end{enumerate} 

A separate section, Section \ref{sec:CCM}, is devoted to the relation between
superintegrable systems in Darboux spaces and two-dimensional spaces of
constant curvature.

\section{Darboux spaces of type two}
\label{sec:Darboux2}

\subsection{The free particle and separating coordinate systems}
\label{subsec:Darboux2free}
If we allow rescaling of the variables $x$ and $y$, as well as the Hamiltonian 
$H$ then we can always take $H$ to be of the form  
\be
\label{H2}
H_0 = {(x-y)^2\over (x-y)^2-1 }\  p_xp_y\,.
\ee
In the coordinates 
$x={1\over 2}(v+iu)$, $y={1\over 2}(v-iu)$ this Hamiltonian becomes
$$
H_0 = \frac{u^2 (p^2_u+p^2_v)}{u^2+1}\,.
$$
Associated with the Hamiltonian are three integrals of the free motion  
$$
K = p_v,\qquad  
X_1 = \frac{2v(p_v^2-u^2p^2_u)}{u^2+1}+2up_up_v,\quad 
X_2 = \frac{(v^2-u^4)p^2_v+u^2(1 -v^2)p^2_u}{u^2+1}+2uvp_up_v.
$$
These three integrals satisfy the following polynomial algebra relations: 
\be
\label{Poissonalgebra2}
\{K,X_1\}=2(K^2-H_0),\qquad 
\{K,X_2\}=X_1,\qquad 
\{X_1,X_2\}=4KX_2.
\ee
They are functionally dependent via the relation 
\be
\label{identity2}
X^2_1 - 4K^2X_2 + 4H_0X_2 - 4H_0^2 = 0\,.
\ee

The corresponding problem in quantum mechanics can  be obtained via the
usual quantisation rules and symmetrisation: 
$$
\hat H_0 = {u^2\over u^2+1 } (\partial ^2_u+\partial ^2_v), \quad
\hat K =\partial _v, \quad 
\hat X_1 = {2v\over (u^2+1 )}(\partial_v^2-u^2\partial ^2_u)
          +2u\partial_u\partial_v + \partial_v,
$$
$$
{\hat X}_2 = 
{1\over u^2+1}\Bigl((v^2-u^4)\partial ^2_v+u^2(1 -v^2)
\partial ^2_u\Bigr)+2uv\partial _u\partial _v+u\partial _u+v\partial
_v-{1\over 4},
$$
where the constant in the last expression is taken for convenience. The 
commutation relations are identical with those of the corresponding 
classical algebra:
$$
[\hat K,\hat X_1] = 2(\hat K^2-\hat H_0)\qquad  
[\hat K,\hat X_2] = \hat X_1,\qquad
[\hat X_1,\hat X_2] = 2\{\hat K,\hat X_2\}.
$$
Here $\{\hat K,\hat X_2\} = \frac12(\hat K\hat X_2+\hat X_2\hat K)$.
The operator relation (that exists in analogy with the functional relation in 
the classical case) is 
$$
\hat X^2_1 - 2\{\hat K^2,\hat X_2\} + 4\hat H_0\hat X_2 - 4 \hat H_0^2-\hat H_0
+4\hat K^2=0.
$$

The line element
$ds^2 = (du^2+dv^2)(u^2+1)/u^2$
can be realised as a two-dimensional surfaced embedded in three dimensions
by
$$
X = \frac{v\sqrt{u^2+1}}u\,,\quad
Y-T = \frac{\sqrt{u^2+1}}u\,, \quad 
Y+T = -\frac{(2u^4+5u^2+8v^2)\sqrt{u^2+1}}{8u} - \frac38\mbox{arcsinh}u\,,
$$
in which case,
$$
ds^2 = dX^2+dY^2-dT^2 = \frac{u^2+1}{u^2}(du^2+dv^2)\,.
$$

We wish to determine all the essentially different separable coordinate
systems for the free classical or quantum particle. 
In order to do this we need to consider a general quadratic constant of the form  
$\lambda=aX_1+bX_2+cK^2$.
Under the adjoint action of $\exp (\alpha K)$, $ X_1$ and $X_2$ transform according
to 
$$
X_1\rightarrow X_1+2\alpha (K^2-H_0),\qquad
X_2\rightarrow X_2+\alpha X_1+\alpha ^2(K^2-H_0).
$$
From these transformation formulae we see that if $b\neq 0$ we can always take 
$\lambda $ in the form $\lambda =X_2+\beta K^2$. If $b=0$ then there are two 
representatives possible: $X_1$ or $K^2$. We  have the following cases:
\be 
\label{freesep2}
X_2+\beta K^2\,,\qquad 
X_1\,,\qquad
K^2\,.
\ee

We now demonstrate the explicit coordinates for each of these 
representatives using methods of our previous article\cite{KKW}.

\subsubsection{Coordinates associated with $X_2+\beta K^2$}
If we choose $\beta =b^2$, $b\neq 0$ suitable 
coordinates $\omega$, $\varphi$ are
\be
\label{coordsD2ell}
u=b\cosh\omega \cos\varphi,\quad v=b\sinh\omega \sin\varphi\,,
\ee
the standard form of elliptical coordinates in the plane. The classical 
Hamiltonian has the form
$$
H_0 = \frac{p^2_\omega +p^2_\varphi}
         {\sec^2\varphi-\sech^2\omega+b^2\left(\cosh^2\omega-\cos^2\varphi\right)}\,.
$$
The corresponding quadratic constant, expressed in these coordinates is  
$$
X_2 + b^2K^2
  = \frac{\left(\sec^2\varphi+b^2\sin^2\varphi\right)p^2_\omega
          +\left(\sech^2\omega-b^2\sinh^2\omega\right)p^2_\varphi}
       {\left(\sec^2\varphi-\sech^2\omega\right)+b^2\left(\cosh^2\omega-\cos^2\varphi\right)}\,.
$$
The Hamilton-Jacobi equation is 
$$
\frac{\pdsq S\omega + \pdsq S\varphi}
     {\sec^2\varphi-\sech^2\omega+b^2\left(\cosh^2\omega-\cos^2\varphi\right)} = E\,,
$$
with solutions of the form
$$
S(\omega,\varphi) = \frac{b\sqrt E}2 \left(
       \int\frac1\Omega\sqrt{\frac{(\Omega+\beta_1)(\Omega+\beta_2)}{\Omega-1}} \ d\Omega
    + \int\frac1\Phi\sqrt{\frac{(\beta_1-\Phi)(\Phi-\beta_2)}{1-\Phi}} \ d\Phi\right)\,.
$$
where $\beta_1+\beta_2=-\lambda/Eb^2$, $\beta_1\beta_2=-1/b^2$, $\Phi=\cos^2\varphi$,
$\Omega=\cosh^2\omega$.
The corresponding Schr\"odinger equation    
$$
\frac{\left(\partial ^2_\varphi +\partial ^2_\omega \right)\Psi}
     {\sec^2\varphi-\sech^2\omega+b^2\left(\cosh^2\omega-\cos^2\varphi\right)}=E\Psi
$$
has solutions of the form  
$$
\Psi =\sqrt{\cos\varphi \cosh\omega}\ S^{m(j)}_n\left(i\sinh\omega,-{1\over 4}Eb\right)
Ps^m_n\left(\sin\varphi ,-{1\over 4}Eb\right)\,,\qquad j=1,2
$$
where $S^{m(j)}_n(z,\kappa )$ and $Ps^m_n(t,\kappa )$ are spheroidal functions 
 \cite{EMOT1} and $E=m^2-{1\over 4}$.

\subsubsection{Coordinates associated with $X_2$}
Here we use polar coordinates:
\be
\label{coordsD2pol}
u=r\cos\theta\,,\qquad v=r\sin\theta\,.
\ee
The classical Hamiltonian has the form
$$
H_0 = \frac{r^2p_r^2+p_\theta^2}{r^2+\sec^2\theta}
$$
and the corresponding quadratic constant is 
$$
X_2 = \frac{r^2\sec^2\theta\, p_r^2-p_\theta^2}{r^2+\sec^2\theta}\,.
$$
The Hamilton-Jacobi equation in these coordinates is
$$
\frac{r^2\pdsq Sr + \pdsq S\theta}
         {r^2+\sec^2\theta} = E\,,
$$
with solution
\beann
S(r,\theta) &=& \sqrt{Er^2+\lambda}
   - \sqrt\lambda \ \mbox{arctanh}\sqrt{\frac{Er^2+\lambda}\lambda} \\
  & & \quad {} - \sqrt\lambda \ \log\left(\sqrt\lambda\sin\theta+\sqrt{E-\lambda\cos^2\theta}\right) \\
  & & \quad {}
   +\frac12\sqrt E \ \mbox{arccosh}\left(\frac{(E+\lambda)\cos^2\theta-2E}
       {(E-\lambda)\cos^2\theta}\right)\,.
\eeann
The corresponding Schr\"odinger equation is   
$$
\frac{\ds r^2\pdtwo\Psi r+\sin\theta\pd{\null}\theta\left(\frac1{\sin\theta}\pd\Psi\theta\right)}
     {r^2+\sec^2\theta} = E\Psi\,,
$$ 
and has solutions of the form  
$$
\Psi =\sqrt{r\sin\theta}\ C_{\ell+\frac12}\left(\sqrt{-E}\ r\right)P^m_\ell(\cos\theta )\,,\qquad E=m^2-{1\over 4}\,, 
$$
where  $C_\nu (z)$ is a Bessel function 
and $P^n_\ell (\cos\theta )$ is an associated Legendre polynomial
\cite{EMOT1}.

\subsubsection{Coordinates associated with $X_1$}
A suitable choice of coordinates is  
\be
\label{coordsD2para}
u=\xi \eta  ,\quad v={1\over 2}(\xi ^2-\eta ^2)\,.
\ee
The classical Hamiltonian in these coordinates has the form 
$$
H_0 = \frac{p^2_\xi +p^2_\eta}{\xi^2+\eta^2+{1\over \xi ^2}+{1\over\eta ^2}}.
$$
The corresponding quadratic constant is
$$
X_1 = \frac{\left(\eta^2+{1\over\eta^2}\right)p^2_\xi
           -\left(\xi^2+{1\over\xi^2}\right)p^2_\eta}
        {\xi^2+\eta^2+{1\over\xi^2}+{1\over\eta^2}}\,.
$$
The Hamilton-Jacobi equation has the form   
$$
{\left({\partial S\over \partial \xi }\right)^2
        +\left({\partial S\over \partial \eta }\right)^2
\over \xi ^2+\eta ^2+{1\over \xi ^2}+{1\over \eta ^2}} = E\,,
$$
which has solution
\beann
S(\xi,\eta) &=&
 - \frac{\sqrt{E\xi^4+E-\lambda\xi^2}}{\xi^2}
 -\frac\lambda{2\sqrt E} \ \mbox{arctanh}\left(
           \frac{\lambda\xi^2-2E}{2\sqrt E\sqrt{E\xi^4+E-\lambda\xi^2}}\right) \\
 & & \quad {}
 + \sqrt E \ \log\left(\sqrt E(2E\xi^2-\lambda)+2E\sqrt{E\xi^4+E-\lambda\xi^2}\right) \\
 & & \quad {}
 - \frac{\sqrt{E\eta^4+E+\lambda\eta^2}}{\eta^2}
 -\frac\lambda{2\sqrt E} \ \mbox{arctanh}\left(
           \frac{\lambda\eta^2+2E}{2\sqrt E\sqrt{E\xi^4+E+\lambda\xi^2}}\right) \\
 & & \quad {}
 + \sqrt E \ \log\left(\sqrt E(2E\xi^2+\lambda)+2E\sqrt{E\xi^4+E+\lambda\xi^2}\right)\,.
\eeann
The corresponding Schr\"odinger equation is    
$$
{\partial ^2_\xi\Psi +\partial ^2_\eta \Psi \over \xi ^2+\eta
^2+{1\over \xi ^2}+{1\over \eta ^2}} =E\Psi\,.
$$
Typical solutions are  
$$
\Psi =\frac1{\sqrt{\xi \eta}}\ M_{\chi ,\mu }(\sqrt{E}\xi ^2)M_{-\chi ,\mu }(
\sqrt{E}\eta ^2)
$$
where $M_{\chi ,\mu }(z)$ is a Whittaker function \cite{BUC} and 
$E=4\mu ^2-{1\over 4}$.

\subsubsection{Coordinates associated with $K^2$}
The representative $K^2$
has associated with it the coordinates $u$ and 
$v$, in which the ignorable variable has a fundamental role to play.  The
Hamiltonian and constant associated with this separation have already
been given.  The Hamilton-Jacobi equation has the form  
$$
{u^2\over u^2+1 } 
\left(\left({\partial S\over \partial u}\right)^2
  +\left({\partial S\over \partial v}\right)^2\right)=E\,,
$$
which has solution, with separation constant $c$,
$$
S(u,v) = \sqrt{u^2(E-c^2)+E} - \sqrt E \ \mbox{arctanh}\sqrt{\frac{u^2(E-c^2)+E}E}+cv\,.
$$
The corresponding Schr\"odinger equation has the form  
$$
{u^2\over u^2+1 } \left(\partial ^2_u\Psi+\partial ^2_v\Psi\right) =E\Psi.
$$
Typical solutions are  
$$
\Psi =\sqrt u\ C_\nu \left(\sqrt{m^2-E}\ u\right)e^{mv}
$$
where $E=\nu ^2-{1\over 4}$.

\bigskip

It is no surprise that the Hamiltonian is separable in
elliptic, parabolic and polar coordinates, since, 
if we write the classical equation $H=E$ in $u,v$ coordinates we obtain 
$$
p^2_u+p^2_v-E\left({1 \over u^2}+1\right)=0.
$$
This equation is essentially the same form as a flat space superintegrable 
system with Cartesian coordinates $u,v$ and potential $\alpha/u^2$, viz
$$
p^2_u+p^2_v+{\alpha \over u^2}-E=0.
$$
It is known to be solvable via the separation of variables Ansatz in
elliptic, Cartesian, polar  and parabolic coordinates. This correspondence 
between flat space superintegrable systems and their curved analogues is 
essentially the way all the curved superintegrable systems can be obtained and is
discussed in more detail in section \ref{sec:CCM}.

\subsection{Superintegrability for Darboux spaces of type two.}
In this section we address the problem of superintegrability for the 
Hamiltonian 
\be
H_0 = \frac{u^2 (p^2_u+p^2_v)}{u^2+1}\,.
\ee
This is done in exactly the same manner as it was for the Darboux space
of type 1 in a previous paper \cite{KKW}. The free space Hamiltonian is given and we
compute the possible potentials that correspond to superintegrability.
There are four possibilities:

\begin{itemize}
\item[{\bf [A]}]
$\ds H = {u^2\over u^2+1 } \left(p^2_u+p^2_v
         +a_1\left({1\over 4}u^2+v^2\right)+a_2v+{a_3\over u^2}\right)$.
\end{itemize}

A basis for the additional constants of the motion is
\beann
R_1 &=& X_1 + \frac{a_1}2 v\left(u^2 + \frac{u^2+4v^2}{u^2+1}\right)
            + \frac{a_2}2\left(u^2+\frac{4v^2}{u^2+1}\right)
            - \frac{2a_3v}{u^2+1}\,, \\
R_2 &=& K^2 + a_1v^2 + a_2v\,.
\eeann
These, along with $R=\{R_1,R_2\}$, form a quadratic algebra
\be
\label{quadalg}
\{R,R_1\} = -\frac12\pd{R^2}{R_2}\,, \qquad \{R,R_2\} = \frac12\pd{R^2}{R_1}
\ee
that is determined by the identity
\beann
R^2 &=& 16R^3_2 - 4a_1R_1^2 - 32HR_2^2 - 8a_2R_1R_2 \\
 & & \quad {} + 8a_2HR_1 + 16(H^2+a_1H-a_1a_3)R_2 + 4a_2^2H - 4a_2^2a_3\,.
\eeann

The classical equation of motion $H-E=0$ is 
$$
p^2_u+p^2_v+a_1\left({1\over 4}u^2+v^2\right)+a_2v+ {a_3-E\over u^2}
-E=0.
$$
The basic form of this equation is  a superintegrable system 
in flat space, but with rearranged constants, which is solvable via separation of
variables in Cartesian and parabolic coordinates.

This accords with the fact that the 
leading part of a quadratic constant for this Hamiltonian will be an element of
the orbits represented by $X_1$ and $K^2$.
So this Hamiltonian also separates in the `parabolic' coordinates $\xi$, $\eta$ 
(\ref{coordsD2para}) and in these coordinates takes the form 
$$
H = \frac{p^2_\xi+p^2_\eta+{1\over 4}a_1\left(\xi ^6+\eta ^6\right)
              +{1\over 2}a_2\left(\xi ^4-\eta ^4\right)
              +a_3\left({1\over \eta ^2}+{1\over \xi ^2}\right)}
         {\xi^2+\eta^2+{1\over\xi^2}+{1\over\eta^2}}\,.
$$

Adding the same potential and coordinate functions to the quantum Hamiltonian $\hat H_0$
and its corresponding commuting operators $\hat X_1$ and $\hat K^2$, we obtain the operators
\beann
\hat H   &=& \hat H_0 + \frac{u^2}{u^2+1}\left(
                  a_1\left({1\over 4}u^2+v^2\right)+a_2v+{a_3\over u^2}\right) \\
\hat R_1 &=& \hat X_1 + \frac{a_1}2 v\left(u^2 + \frac{u^2+4v^2}{u^2+1}\right)
            + \frac{a_2}2\left(u^2+\frac{4v^2}{u^2+1}\right)
            - \frac{2a_3v}{u^2+1}\,, \\
\hat R_2 &=& \hat K^2 + a_1v^2 + a_2v\,.
\eeann
$\hat R_1$ and $\hat R_2$ commute with $\hat H$ and along with
$\hat R = [\hat R_1,\hat R_2]$, obey the 
corresponding quantum quadratic algebra relations
\beann
{}[\hat R,\hat R_1]
  &=& -24\hat R^2_2 + 4a_2\hat R_1 + 32\hat H\hat R_2
      - 8\hat H^2 - 8a_1\hat H + 6a_1 + 8a_1a_3\,, \\
{}[\hat R,\hat R_2]
  &=& -4a_1\hat R_1 - 4a_2\hat R_2 + 4a_2\hat H
\eeann
and the operator identity
\beann
\hat R^2 &=& 16\hat R^3_2 
- 4a_1\hat R^2_1 
- 32\hat H\hat R^2_2 
- 4a_2\{\hat R_1,\hat R_2\} \\
& & \quad {} + 8a_2\hat H\hat R_1 
+ 16\hat H^2\hat R_2 + 16a_1\hat H\hat R_2 - 4a_1(4a_3-11)\hat R_2 \\
& & \quad {} + 4\left(a_2^2+8a_1\right)\hat H
- 4b^2_2\left(a_3+\sfrac34\right)\,.
\eeann

\begin{itemize}
\item[{\bf [B]}]
$\ds H = {u^2\over u^2+1}\left(p^2_u+p^2_v+b_1\left(u^2+v^2\right)+{b_2\over u^2}+
{b_3\over v^2}\right)$.
\end{itemize}

The additional constants of the motion have the form 
$$
R_1 = X_2 + \frac{u^2+v^2}{u^2+1}\left(b_1(u^2+v^2)
                    - b_2 - b_3\frac{u^2}{v^2}\right)\,,\qquad 
R_2 = K^2 + b_1v^2 + {b_3\over v^2}\,.
$$
The corresponding quadratic algebra relations can be determined, using
(\ref{quadalg}), from the identity 
\beann
R^2 &=& 16R_1R_2^2 
- 16b_1R_1^2 
- 16HR_1R_2 
+ 32b_1(H-b_2-b_3)R_1 
+ 16(H+b_3-b_2)HR_2 \\
 & & \quad {}
- 16(b_1+b_3)H^2
+ 32b_1(b_2-b_3)H 
- 16b_1(b_2-b_3)^2\,.
\eeann

The equation of motion $H-E=0$ becomes
$$
p^2_u+p^2_v+b_1(u^2+v^2)+ {(b_2-E)\over u^2} + {b_3\over v^2}
-E=0\,.
$$
This is a superintegrable system 
in flat space, but with rearranged constants, which is solvable via separation of
variables in Cartesian, polar and elliptic coordinates.  Again, this agrees with
the observation that for this Hamiltonian we have quadratic constants
with leading parts $K^2$, $X_2$ and $X_2+\beta K^2$.
In the latter two coordinate systems, the Hamiltonian takes the forms:
\begin{itemize}
\item[(i)] Elliptical coordinates (\ref{coordsD2ell}) 
$$
H = \frac{p^2_\omega+p^2_\varphi + \sfrac14b_1b^2(\sinh^22\omega + \sin^22\varphi)
           + b_2(\sec^2\varphi - \sech^2\omega)
           + b_3(\cosec^2\varphi + \cosech^2\omega)}
   {\sec^2\varphi-\sech^2\omega + b^2\left(\cosh^2\omega-\cos^2\varphi\right)}\,.
$$
\item[(ii)]
Polar coordinates (\ref{coordsD2pol})
$$
H= \frac{r^2p^2_r+p^2_\theta + b_1r^4 + b_2\sec^2\theta + b_3\cosec^2\theta}
        {r^2+\sec^2\theta}\,.
$$
\end{itemize}

The corresponding quantum algebra relations are 
\beann
[\hat R,\hat R_1]
  &=& - 8\{\hat R_1,\hat R_2\} + 8\hat H\hat R_1 + 12\hat R_2
               - 8\hat H^2 + 8(b_2-b_3-\sfrac34)\hat H \,,\\
{}[\hat R,\hat R_2]
  &=&  8\hat R^2_2 - 16b_1\hat R_1 - 8\hat H\hat R_2 + 16b_1\hat H
                - 16b_1(b_2+b_3+\sfrac34)\,, \\
\hat R^2
  &=& 8\{\hat R_1,\hat R_2^2\}
- 8\hat H\{\hat R_1,\hat R_2\}
+ 16\hat H^2\hat R_2
- 16b_1\hat R_1^2
- 76\hat R_2^2 \\
& & \ {}
+ 32b_1\hat H\hat R_1
- 8b_1(4(b_3+b_2)+3)\hat R_1
+ 16(b_3-b_2+\sfrac{19}4)\hat H\hat R_2 \\
& & \ {}
- 16(b_1+b_3+\sfrac34)\hat H^2
- 8b_1(4(b_3-b_2)+3)\hat H
+ b_1\Bigl(36+48b_3-(4(b_3-b_2)+3)^2\Bigr)
\eeann

\begin{itemize}
\item[{\bf [C]}]
$\ds H=\frac{p^2_\xi+p^2_\eta+c_1+{c_2\over \xi ^2}+{c_3\over \eta ^2}}
{\xi^2+\eta^2+{1\over\xi^2}+{1\over\eta^2}}\,.
$
\end{itemize}

The additional constants of the motion are 
\beann
R_1 &=& X_1 + \frac{c_1\xi^2(\eta ^4+1)+c_2(\eta^4+1)-c_3(\xi^4+1)}
                   {(\xi^2\eta^2+1)(\xi^2+\eta^2)}\,, \\
R_2 &=& X_2 + \frac{c_1(\xi^2+\eta^2)-c_2(\eta^4-1)-c_3(\xi^4-1)}
                   {4(\xi^2\eta^2+1)}\,.
\eeann
The corresponding Poisson algebra can be determined from the
identity 
\beann
R^2 &=& 4R_1^2R_2
- (c_2+c_3)R_1^2
+ 16HR_2^2
- 4c_1R_1R_2
+ 2c_1c_3R_1
- 16H^2R_2 \\
& & \quad {}
+ 4(c_2+c_3)H^2
+ (c_1^2-4c_2c_3)H
- c_1^2c_3
\eeann
The Hamiltonian can be written in separable form for the following coordinate 
systems.
\begin{itemize}
\item[(i)] Displaced Elliptic coordinates \ $\xi=b'\cosh\omega'\cos\varphi'$,  
$\eta =b'\sinh\omega'\sin\varphi'$.
$$
H= \frac{p^2_{\omega'} +p^2_{\varphi'}
             + c_1b^{\prime2}(\cosh^2\omega' -\cos ^2\varphi' )
             + c_2(\sec^2\varphi'-\sech^2\omega')
             + c_3(\cosec^2\varphi'+\cosech^2\omega')}
        {b^{\prime4}(\cosh^4\omega'-\cos^4\varphi'-\cosh^2\omega'+\cos^2\varphi')
           +\sec^2\varphi'+\cosec^2\varphi' + \cosech^2\omega' - \sech^2\omega'}\,.
$$
These coordinates are not those given in (\ref{coordsD2ell}) and are related to
$u$ and $v$ by
$$
u = \frac14b^{\prime2}\sinh2\omega'\sin2\varphi'\,,\qquad
v = \frac14b^{\prime2}\left(\cosh2\omega'\cos2\varphi'+1\right)\,.
$$
\item[(ii)] Polar coordinates \ $\xi =r'\cos\theta'$, $\eta =r'\sin\theta' $. 
$$
H = \frac{r'^2p^2_{r'}+p^2_{\theta'}+c_1r'^2+c_2\cosec^2\theta'+c_3\sec^2\theta'}
         {r'^4+\sec^2\theta'+\cosec^2\theta'}\,.
$$
These coordinates are not those given in (\ref{coordsD2pol}) and are related to
$u$ and $v$ by
$$
u = \frac12r'^2\sin2\theta'\,, \qquad v = \frac12r'^2\cos2\theta'\,.
$$
\end{itemize}

The corresponding quantum algebra relations are
\beann
[\hat R,\hat R_1]
  &=& - 2\hat R^2_1 - 2c_1\hat R_1 - 16\hat H\hat R_2 + 8\hat H^2 - 6\hat H \\
{}[\hat R,\hat R_2]
  &=& 2\{\hat R_1,\hat R_2\} - (c_2+c_3)\hat R_1 - 2c_1\hat R_2 + c_1c_3 \\
\hat R^2
  &=& 2\{\hat R_1^2,\hat R_2\}
+ 16\hat H\hat R_2^2
- (c_2+c_3+4)\hat R_1^2
- 2c_1\{\hat R_1,\hat R_2\}
+ 2c_1(c_3+2)\hat R_1 \\
 & & \ {}
- 16\hat H^2\hat R_2
+ 12\hat H\hat R_2
+ 4(c_2+c_3)\hat H^2
+ (c_1^2-4c_2c_3-3(c_2+c_3))\hat H
- \sfrac14(3+4c_3)c_1^2
\eeann

The equation of 
motion $H-E=0$  is
$$
p^2_\xi +p^2_\eta +c_1-E(\xi ^2+\eta ^2)+ {(c_2-E)\over \xi ^2} + 
{(c_3-E)\over \eta ^2} =0.
$$
This is  a superintegrable system 
in flat space, but with rearranged constants, which is solvable via separation of
variables in Cartesian, polar and elliptic coordinates.

\begin{itemize}
\item[{\bf [D]}]
$\ds H = {u^2(p^2_u+p^2_v+d)\over u^2+1}$.
\end{itemize}

The additional constants of the motion are 
$$
R_1 = X_1 + {2dv\over u^2+1},\qquad
R_2 = X_2 + \frac{d(u^2+v^2)}{u^2+1}, \qquad
K=p_v.
$$
The corresponding Poisson algebra relations are 
$$
\{K,R_1\}=2K^2-2H+2d\,,\qquad  
\{K,R_2\}=R_1\,,\qquad 
\{R_1,R_2\}=-4KR_2\,.
$$
The functional relation between these constants is
$$
R^2_1 - 4K^2R_2 + 4(H-d)R_2 - 4H^2 + 4dH = 0\,.
$$
The Hamiltonian can be written in separable form for all the possible types of 
separable coordinates we have discussed viz.
\begin{itemize}
\item[(i)] Elliptic coordinates (\ref{coordsD2ell}).
$$
H = \frac{p^2_\omega+p^2_\varphi+b^2d(\cosh^2\omega-\cos^2\varphi)}
              {b^2(\cosh^2\omega-\cos^2\varphi)+\sec^2\varphi-\sech^2\omega}\,.
$$
\item[(ii)] Polar coordinates (\ref{coordsD2pol}).
$$
H = \frac {r^2p^2_r+p^2_\theta +dr^2}
          {r^2+\sec^2\theta}\,.
$$
\item[(iii)] Parabolic coordinates (\ref{coordsD2para}).
$$
H = \frac{p^2_\xi+p^2_\eta+d(\xi^2+\eta^2)}
         {\xi^2+\eta^2+{1\over \xi ^2}+{1\over \eta ^2}}\,.
$$
\end{itemize}

The corresponding quantum algebra relations have the form 
$$
[\hat K,\hat R_1]=2\hat K^2-2\hat H+2d,\qquad 
[\hat K,\hat R_2]=\hat R_1,\qquad 
[\hat R_1,\hat R_2]=2\{\hat K,\hat R_2\}.
$$
The operator identity satisfied by the defining operators of the 
quantum algebra is
$$
\hat R^2_1-2\{\hat K^2,\hat R_2\}
+4\hat H\hat R_2
-4d\hat R_2
+4\hat K^2
-4\hat H^2
+(4d-1)\hat H=0\,.
$$
The equation of 
motion $H-E=0$ is 
$$
p^2_u+p^2_v+d-E-{E\over u^2} =0.
$$
This is  a superintegrable system 
in flat space, but with rearranged constants, which is solvable via separation of
variables in Cartesian, polar, elliptic and parabolic coordinates.

\section{Darboux spaces of type three.}
\label{sec:Darboux3}

\subsection{The free particle and separating coordinate systems}
\label{subsec:Darboux3free}
With rescaling and translation of the variables $x$ and $y$ the 
Hamiltonian $H$  has the form 
\be
H_0 = \frac{e^{(x+y)/2}}{1+e^{-(x+y)/2}}\, p_xp_y.
\ee
In coordinates $x=u-iv$, $y=u+iv$ we can  write 
this Hamiltonian in positive definite form 
$$
H_0 ={1\over 4} {e^{2u}(p^2_u+p^2_v)\over e^u+1}.
$$
Associated with the Hamiltonian are three integrals of the free motion 
$$
K = p_v\,, \qquad  
X_1 = {1\over 4} {e^{2u}\over e^u+1} \cos v\, p^2_u
      - {1\over 4}{e^u(e^u+2)\over e^u+1} \cos v\, p^2_v
      + {1\over 2} e^u \sin v\, p_up_v\,,
$$
$$
X_2 = {1\over 4} {e^{2u}\over e^u+1} \sin v\, p^2_u
     - {1\over 4} {e^u(e^u+2)\over e^u+1} \sin v\, p^2_v
     - {1\over 2} e^u \cos v\, p_up_v\,.
$$
The  integrals satisfy the polynomial algebra relations 
$$
\{K,X_1\}=-X_2,\qquad \{K,X_2\}=X_1,\qquad  
\{X_1,X_2\}=KH_0.
$$
They are functionally dependent via the relation 
$$
X^2_1+X^2_2-H_0^2-H_0K^2=0.
$$

The corresponding problem in quantum mechanics can readily be obtained via 
the usual quantisation rules and symmetrisation.  
$$
\hat H_0 = \frac14 \frac{e^{2u}}{e^u+1}\left(\partial_u^2+\partial_v^2\right)\,,
\qquad \qquad
\hat K   = \partial_v\,,
$$
\beann
\hat X_1 &=& \frac14\frac{e^{2u}}{e^u+1}\cos v\partial_u^2
               - \frac14\frac{e^u(e^u+2)}{e^u+1}\cos v\partial_v^2
               + \frac12e^u\sin v\partial_u\partial_v 
               + \frac14e^u\cos v\partial_u + \frac14e^u\sin v\partial_v\,, \\
\hat X_2 &=& \frac14\frac{e^{2u}}{e^u+1}\sin v\partial_u^2
               - \frac14\frac{e^u(e^u+2)}{e^u+1}\sin v\partial_v^2
               - \frac12e^u\cos v\partial_u\partial_v 
               + \frac14e^u\sin v\partial_u - \frac14e^u\cos v\partial_v\,. \\
\eeann
The commutator algebra obtained
has the same form as the Poisson algebra, and the identity relating the operators
is
$$
\hat X^2_1+\hat X^2_2-\hat H_0^2-\hat H_0\hat K^2+\frac14\hat H_0=0\,.
$$

The line element
$ds^2 = (e^{-u}+e^{-2u})(du^2+dv^2)$
can be realised as a two-dimensional surface embedded in three dimensions
by
$$
X = v\sqrt{e^{-u}+e^{-2u}}\,, \qquad
Y-T = \sqrt{e^{-u}+e^{-2u}}
$$
$$
Y+T = (1-v^2)\sqrt{e^{-u}+e^{-2u}} + \log\Big(1+2e^{-u}+2\sqrt{e^{-u}+e^{-2u}}\Bigr)
         + \frac12\mbox{arctan}\Bigl(2\sqrt{e^{-u}+e^{-2u}}\Bigr)\,,
$$
in which case,
$$
ds^2 = dX^2+dY^2-dT^2 = (e^{-u}+e^{-2u})(du^2+dv^2)\,.
$$

Just as we have done in other cases, we wish to determine all the essentially 
different separable coordinate systems for the free classical or quantum 
particle.  To do this we need to consider a general quadratic constant of the 
form 
$
\lambda =aX_1+bX_2+cK^2$.
Under the adjoint action of $\exp (\alpha K)$, $ X_1$ and $X_2$ transform according
to 
$$
X_1\rightarrow  \cos\alpha \ X_1-\sin\alpha \ X_2,\qquad 
X_2\rightarrow  \sin\alpha \ X_1+\cos\alpha \ X_2.
$$
From this transformation law we see that $\lambda$ can take five different 
forms:
\be
K^2,\qquad 
X_1,\qquad
X_1+\gamma K^2,\qquad
X_1+iX_2,\qquad 
X_1+iX_2-K^2.
\label{freesep3}
\ee

We now demonstrate the explicit coordinates in the case of each of these 
representatives. 

\subsubsection{Coordinates associated with $K^2$}
These are the coordinates associated with the ignorable coordinate $v$ and the
Hamiltonian has already been given in the $u,v$ coordinates.  
The Hamilton-Jacobi equation is
$$
{1\over 4} {e^{2u}\over e^u+1} 
\left(\pdsq Su+\pdsq Sv\right) = E\,,
$$
with solutions
\beann
S(u,v) &=& - \frac{\sqrt{4E(1+e^u)-c^2e^{2u}}}{ce^u}
 - \frac{\sqrt E}c \mbox{arctanh}\left(\frac{\sqrt E(e^u+2)}{\sqrt{4E(1+e^u)-c^2e^{2u}}}\right) \\
 & & \quad {}
   + i\log\left(i(c^2e^u-2E)+c\sqrt{4E(1+e^u)-c^2e^{2u}}\right) + cv\,.
\eeann
The corresponding Schr\"odinger equation is 
$$
{1\over 4} {e^{2u}\over e^u+1} (\partial ^2_u+\partial ^2_v)\Psi
=E\Psi,
$$
with solutions of the form 
$$
\Psi=e^{-u/2} M_{-1/\sqrt{-E},\,\pm m}(4\sqrt{-E}\, e^{-u})e^{imv}.
$$

\subsubsection{Coordinates associated with $X_1$}
For the second representative in (\ref{freesep3}), a suitable
choice of variables is 
\be
\label{coordsD3X1}
\xi =2e^{-u/2}\cos{v\over 2}\,, \qquad
\eta =2e^{-u/2}\sin{v\over 2}\,.
\ee
In terms of these coordinates the classical Hamiltonian has the form 
$$
H_0 = \frac{p^2_\xi +p^2_\eta}
           {4+\xi ^2+\eta ^2}
$$
and the corresponding quadratic constant  is 
$$
X_1 = \frac{(2+\eta ^2)p^2_\xi-(2+\xi^2)p^2_\eta}
         {2(4+\xi ^2+\eta ^2)}\,.
$$
In $\xi$, $\eta$ coordinates the classical Hamilton-Jacobi equation is 
$$
\frac{\pdsq S\xi + \pdsq S\eta}{4+\xi^2+\eta^2} = E\,,
$$
which has the solution
\beann
S &=& \frac12\xi\sqrt{E\xi^2+2E-\lambda}
      + \left(\frac{2E-\lambda}{2\sqrt E}\right)
               \log\left(E+\sqrt{E\xi^2+2E-\lambda}\right) \\
 & & \ {} + \frac12\eta\sqrt{E\eta^2+2E+\lambda}
            + \left(\frac{2E+\lambda}{2\sqrt E}\right)
                    \log\left(E+\sqrt{E\eta^2+2E+\lambda}\right)\,.
\eeann
The Schr\"odinger equation is 
$$
\frac{(\partial ^2_\xi +\partial ^2_\eta )\Psi}
     {4+\xi^2+\eta^2} = E\Psi\,,
$$
which has typical solutions 
$$
\Psi = D_{(\lambda-2E)/\sqrt{4E}}\Bigl(\pm(4E)^{1/4}\xi\Bigr) 
       D_{-(\lambda+E)/\sqrt{4E}}\Bigl(\pm(4E)^{1/4}\eta\Bigr)\,,
$$
in terms of parabolic cylinder functions $D_\nu (z)$ \cite{EMOT1, BUC}.

\subsubsection{Coordinates associated with $X_1+\gamma K^2$}
For the third case it is convenient to take the representative as 
$b^2X_1+2K^2$. Here we identify coordinates via  
\be
\label{coordsD3X1K}
\xi = b\cosh\omega\cos\varphi\,, \qquad 
\eta = b\sinh\omega\sin\varphi\,.
\ee
The classical Hamiltonian  has the form 
$$
H_0 = \frac{p_\omega^2+p_\varphi^2}
           {2b^2(\cosh2\omega-\cos2\varphi)
            + \frac14b^4(\cosh^22\omega-\cos^22\varphi)}
$$
and the corresponding quadratic constant in these coordinates is
$$
b^2X_1+2K^2 =
   \frac{(8\cos2\varphi-b^2\sin2\varphi)p_\omega^2
           + (8\cosh2\omega+b^2\sinh2\omega)p_\varphi^2}
        {8b^2(\cosh2\omega-\cos2\varphi)
            + b^4(\cosh^22\omega-\cos^22\varphi)}\,.
$$
In the $\varphi$, $\omega$ coordinates the classical Hamilton-Jacobi 
equation has the form 
$$
\frac{\pdsq S\omega + \pdsq S\varphi}
         {2b^2(\cosh2\omega-\cos2\varphi)
            + \frac14b^4(\cosh^22\omega-\cos^22\varphi)} = E\,,
$$
and  has the solution
$$
S(\omega,\varphi) = \frac14b^2\sqrt E\left(
  \int\sqrt{\frac{(\Omega-\alpha_1)(\Omega-\alpha_2)}{\Omega^2-1}} \ d\Omega
  + \int\sqrt{\frac{(\beta_1-\Phi)(\Phi-\beta_2)}{1-\Phi^2}} \ d\Phi\right)
$$
where $\alpha_1+\alpha_2=-\beta_1-\beta_2=-8/b^2$, 
$\alpha_1\alpha_2=-\beta_1\beta_2-4\lambda/Eb^2$,
$\Omega=\cosh2\omega$, $\Phi=\cos2\varphi$.
The corresponding Schr\"odinger equation 
$$
\frac{\partial_\omega^2\Psi + \partial_\varphi^2\Psi}
         {2b^2(\cosh2\omega-\cos2\varphi)
            + \frac14b^4(\cosh^22\omega-\cos^22\varphi)} = E\Psi\,,
$$
separates with $\Psi = \Phi(\varphi)\Omega(\omega)$ in the equations
\beann
\Bigl(\partial_\varphi^2+2b^2E\cos2\varphi+\sfrac18b^4E\cos4\varphi+\lambda
          + \sfrac18b^4E\Bigr)\Phi &=& 0\,, \\
\Bigl(\partial_\omega^2-2b^2E\cos2\omega-\sfrac18b^4E\cos4\omega-\lambda
          - \sfrac18b^4E\Bigr)\Omega &=& 0\,, \\
\eeann
which has typical solutions
\beann
\Psi_1 &=& \mbox{gc}_m\left(\varphi,b\sqrt{-E},2b\sqrt{-E}\right)
           \mbox{gc}_m\left(i\omega,b\sqrt{-E},2b\sqrt{-E}\right) \\
\Psi_2 &=& \mbox{gs}_m\left(\varphi,b\sqrt{-E},2b\sqrt{-E}\right)
           \mbox{gs}_m\left(i\omega,b\sqrt{-E},2b\sqrt{-E}\right) \\
\eeann
with corresponding separation constant given by $\lambda_m=\mu_mb^2(1+b^2)E/8$.  The functions appearing here are
even and odd Whittaker Hill functions \cite{MIL}.

\subsubsection{Coordinates associated with $X_1+iX_2$}
In the case of a system specified by the fourth representative 
there are, in fact, no separable coordinates.  However, in the
coordinates
$$
x=\xi+i\eta\,, \qquad y=\frac12\left(\xi-i\eta\right)^2\,,
$$
the classical Hamiltonian takes the form
$$
H_0 = \frac{2p_xp_y}{2y^{-1/2}+x}
$$
and the corresponding constant is
$$
X_1+iX_2=2yH_0-p_x^2\,.
$$
The solution of the Hamilton-Jacobi equation
$$
\frac{2\pd Sx\pd Sy}{2y^{-1/2}+x} = E
$$
is
$$
S = x\sqrt{Ey-\lambda} + \sqrt E\log\left(\sqrt E\ y-\frac\lambda{2\sqrt E}
                 +\sqrt{Ey^2-\lambda y}\right)\,.
$$
The corresponding Schr\"odinger equation is
$$
\frac{2\partial_x\partial_y\Psi}{2y^{-1/2}+x} = E\Psi
$$
which has solutions
$$
\Psi = \frac{\left(2E^{3/2}y-E^{1/2}+2E\sqrt{Ey^2-\lambda y}\right)^{\sqrt E}
                e^{x\sqrt{Ey-\lambda}}}
            {\sqrt{Ey-\lambda}}\,.
$$

\subsubsection{Coordinates associated with $X_1+iX_2-K^2$}
In the case of a system specified by the fifth representative an 
appropriate choice of coordinates is  
\be
\label{coordsD3X1X2K}
\xi = \frac{\mu-\nu}{2\sqrt{\mu\nu}} +\sqrt{\mu\nu}\,,\qquad
\eta =i\left(\frac{\mu-\nu}{2\sqrt{\mu\nu}}-\sqrt{\mu\nu}\right)\,.
\ee
The corresponding classical Hamiltonian has the form 
$$
H_0 = \frac{\mu^2p^2_\mu-\nu^2p^2_\nu}
           {(\mu+\nu)(2+\mu-\nu)},
$$
and the quadratic constant  is 
$$
X_1+iX_2-K^2 = \frac{\nu^2(\mu+2)\mu p^2_\nu - \mu^2(\nu-2)\nu p^2_\mu}
                    {(\mu +\nu )(2+\mu -\nu)}\,.
$$
In the $\mu$, $\nu$ coordinate system the classical Hamilton-Jacobi 
equation has the form
$$
\frac{\mu^2\pdsq S\mu - \nu^2\pdsq S\nu}
        {(\mu +\nu )(2+\mu-\nu)} = E\,,
$$
which has the solution
\beann
S(\mu,\nu) &=&
     \sqrt{E\mu^2+2E\mu+\lambda}
   + \sqrt E \ \log\left(\sqrt E(1+\mu) + \sqrt{E\mu^2+2E\mu+\lambda}\right) \\
 & & \qquad {}
   - \sqrt\lambda \ \mbox{arctanh}\left(\frac{\lambda+E\mu}
                                             {\sqrt\lambda\sqrt{E\mu^2+2E\mu+\lambda}}\right) \\
 & & \ {}
   + \sqrt{E\nu^2-2E\nu+\lambda}
   + \sqrt E \ \log\left(\sqrt E(1-\nu)+\sqrt{E\nu^2-2E\nu+\lambda}\right) \\
 & & \qquad {}
   - \sqrt\lambda \ \mbox{arctanh}\left(\frac{\lambda-E\nu}
                                             {\sqrt\lambda\sqrt{E\nu^2-2E\nu+\lambda}}\right)\,.
\eeann
The Schr\"odinger equation 
$$
\frac{\mu\partial_\mu\left(\mu\partial_\mu\Psi\right)
        - \nu\partial_\nu\left(\nu\partial_\nu\Psi\right)}
     {(\mu+\nu)(2+\mu-\nu)}
 = E\Psi\,,
$$
separates with $\Psi=A(\mu)B(\nu)$ into the equations
$$
\Bigl(\mu\partial_\mu(\mu\partial_\mu)-E\mu^2-2E\mu-\rho^2\Bigr)A(\mu) = 0\,, \qquad
\Bigl(\nu\partial_\nu(\nu\partial_\nu)-E\nu^2-2E\nu-\rho^2\Bigr)B(\nu) = 0\,, 
$$
and has solutions, in terms of the Whittaker function $M_{\lambda,\chi}$, of the form \cite{BUC}
$$
\frac1{\sqrt{\mu\nu}}M_{\sqrt E,\rho}\left(2\sqrt E\ \mu\right)
                     M_{-\sqrt E,\rho}\left(2\sqrt E\ \nu\right).
$$
If we 
write the classical equation $H=E$ in $\xi$, $\eta $ coordinates we obtain 
$$
p^2_\xi +p^2_\eta -E(4+\xi ^2+\eta ^2)=0.
$$
This is in the form of a flat space superintegrable system which can be solved 
by separation of variables in Cartesian, polar, hyperbolic and elliptic 
coordinates. 

\subsection{Superintegrability for Darboux spaces of type three.}
\label{subsec:superint3}
In this section we address the problem of superintegrability for the 
Hamiltonian 
\be
H={1\over 4} {e^{2u}(p^2_u+p^2_v)\over e^u+1}.
\ee
We arrive at three possibilities: [{\bf A}],  [{\bf B}], [{\bf C}].

\begin{itemize}
\item[{\bf [A]}]
$\ds H= {p^2_\xi +p^2_\eta +a_1\xi +a_2\eta +a_3\over 4+\xi ^2+\eta ^2}$
\end{itemize}
The additional constants have the form
\beann
R_1 &=& X_1 + \frac{2a_1\xi(2+\eta^2)-2a_2\eta(2+\xi^2)+a_3(\eta^2-\xi^2)}
                   {4(4+\xi^2+\eta^2)}\,, \\
R_2 &=& X_2 + \frac{a_1\eta(\eta^2-\xi^2+4)+a_2\xi(\xi^2-\eta^2+4)-2a_3\xi\eta} 
                   {4(4+\xi^2+\eta^2)}\,.
\eeann
The corresponding quadratic algebra can be determined from the identity
\beann
R^2
 &=& HR_1^2 + HR_2^2
       + \sfrac18(a_2^2-a_1^2)R_1 - \sfrac14a_1a_2R_2 \\
 & & \ {}
 - H^3 + \sfrac12a_3H^2 + \sfrac1{16}(2a_2^2+2a_1^2-a_3^2)H
 - \sfrac1{32}a_3(a_1^2+a_2^2)\,.
\eeann
This
Hamiltonian separates in a family of coordinate systems obtained by
translating the given separable system via  $\xi\to \xi + a$, \ $\eta\to \eta-a$.
The corresponding quantum algebra relations are
\beann
[\hat R,\hat R_1]
  &=& - \hat H\hat R_2 + \sfrac18a_1a_2\,, \qquad
{}[\hat R,\hat R_2]
  = \hat H\hat R_1 + \sfrac1{16}(a_2^2-a_1^2)\,, \\
\hat R^2
  &=& \hat H\hat R_1^2 + \hat H\hat R_2^2
         + \sfrac18(a_2^2-a_1^2)\hat R_1 - \sfrac14a_1a_2\hat R_2 \\
  & & \ {} - \hat H^3 + \sfrac12(a_3+\sfrac12)\hat H^2 
                + \sfrac1{16}(2a_1^2+2a_2^2-a_3^2)\hat H - \sfrac1{32}a_3(a_1^2+a_2^2)
\eeann
As in the case of free motion, the equation 
$H=E$ becomes
$$
p^2_\xi +p^2_\eta +a_1\xi +a_2\eta +a_3-E(4+\xi ^2+\eta ^2)=0\,.
$$
Again, this is  a superintegrable system in flat space but with rearranged constants.

\begin{itemize}
\item[{\bf [B]}]
$\ds H = \frac{p^2_\xi +p^2_\eta + \frac{b_1}{\xi^2} +  \frac{b_2}{\eta^2}+b_3}
              {4+\xi^2+\eta^2}\,.$
\end{itemize}
The additional constants are
\bea
R_1 &=& X_1 + \frac{2b_1\eta^2(\eta^2+2)-2b_2\xi^2(\xi^2+2)+b_3(\eta^2-\xi^2)}
                   {4(4+\xi^2+\eta^2)}\,, \\
R_2 &=& K^2 + \frac{b_1\eta^2}{4\xi^2} + \frac{b_2\xi^2}{4\eta^2}\,.
\eea
The corresponding quadratic algebra relations are determined by 
\beann
R^2 &=&
- 4R_1^2R_2
- (b_1+b_2)R_1^2
+ 4HR_2^2
+ 2(b_1-b_2)HR_1
+ \sfrac12b_3(b_2-b_1)R_1 \\
 & & \ {}
+ 4H^2R_2
- 2b_3HR_2
+ \sfrac14b_3^2R_2
- (b_1+b_2)H^2 \\
 & & \ {} 
+ \left(\sfrac12b_3(b_1+b_2)-b_1b_2\right)H
- \sfrac1{16}b_3^2(b_1+b_2).
\eeann
This Hamiltonian separates in all the separable coordinate systems given 
in Section \ref{subsec:Darboux2free}.  The Hamiltonian has the explicit forms 

\begin{itemize}
\item[(i)]
In $u$, $v$ coordinates,
$$
H = \frac{e^{2u}\left(p^2_u+p^2_v
      +\frac14b_1\sec^2\frac v2 + \frac14b_2\cosec^2\frac v2 + b_3e^{-u}\right)}
         {4(e^u+1)}\,.
$$
\item[(ii)]
In the elliptical coordinates (\ref{coordsD3X1K}),
$$
H = \frac{p_\omega^2+p_\varphi^2 + b_1\left(\sec^2\varphi-\sech^2\omega\right)
            + b_2\left(\cosec^2\varphi+\cosech^2\omega\right)
            + b_3b^2\left(\cosh^2\omega-\cos^2\varphi\right)}
         {2b^2(\cosh2\omega-\cos2\varphi)
            + \frac14b^4(\cosh^22\omega-\cos^22\varphi)}\,.
$$
\end{itemize}
The corresponding quantum algebra relations have the form 
\beann
[\hat R,\hat R_1]
 &=& 2\hat R_1^2 - 4\hat H\hat R_2 - 2\hat H^2 + \left(b_3+\sfrac12\right)\hat H -\sfrac18b_3^2 \\
{}[\hat R,\hat R_2]
 &=& - 2\{\hat R_1,\hat R_2\} - (b_1+b_2+1)\hat R_1 + (b_1-b_2)\hat H
          + \sfrac14(b_2-b_1)b_3\,, \\
\hat R^2
 &=& -2\{\hat R_1^2,\hat R_2\}
- (b_1+b_2+5)\hat R_1^2
+ 4\hat H\hat R_2^2 \\
 & & \ {}
+ 2(b_1-b_2)\hat H\hat R_1
+ b_3(b_2-b_1)\hat R_1
+ 4\hat H^2\hat R_2
- (2b_3-1)\hat H\hat R_2
+ \sfrac14b_3^2\hat R_2 \\
 & & \ {}
- (b_1+b_2-2)\hat H^2
+ \left(\sfrac12(b_3+\sfrac32)(b_1+b_2)-b_3-b_1b_2-\sfrac12\right)H
- \sfrac1{16}b_3^2(b_1+b_2-2).
\eeann
As in the case of free motion, we observe that  equation 
 $H=E$ becomes
$$
p^2_\xi +p^2_\eta + {b_1\over \xi ^2} + {b_2\over \eta ^2} 
+b_3-E(4+\xi ^2+\eta ^2)=0.
$$
This is  a superintegrable system 
in flat space, with rearranged constants, that  separates variables in 
Cartesian, polar and elliptic coordinates.

\begin{itemize}
\item[{\bf [C]}]
$\ds
H = \frac{\mu^2p^2_\mu-\nu^2p^2_\nu
          +c_1(\mu +\nu )+c_2\frac{\mu+\nu}{\mu\nu}+c_3\frac{\mu^2-\nu^2}{\mu^2\nu^2}}
        {(\mu+\nu)(2+\mu-\nu)}$.
\end{itemize}
The additional constants of the motion have the form 
$$
R_1 =X_1 + iX_2 - \frac{c_1\mu^2\nu^2+c_2\mu\nu+2c_3(1+\mu-\nu)}
                          {\mu\nu(2+\mu-\nu)} \qquad
R_2 = K^2 - c_2\frac{\mu-\nu}{\mu\nu} - c_3\frac{(\mu-\nu)^2}{\mu^2\nu^2}\,.
$$
The corresponding quadratic Poisson algebra relations can be determined from
\beann
R^2 &=& 
-4R_2R_1^2
+8c_2HR_1
-4c_1c_2R_1
+16c_3HR_2 \\
 & & \ {}
+16c_3H^2
+4(c_2^2-4c_1c_3)H
+4c_1^2c_3\,.
\eeann

The quantum algebra relations are 
$$
[\hat R,\hat R_1] = 2\hat R_1^2 - 8c_3\hat H\,,\qquad 
[\hat R,\hat R_2]
  =  -2\{\hat R_1,\hat R_2\} - \hat R_1 + 4c_2\hat H - 2c_1c_2\,,
$$
$$
\hat R^2 = -2\{\hat R_1^2,\hat R_2\} + 8c_2\hat H\hat R_1 + 16c_3\hat H\hat R_2
              - 5\hat R_1^2 - 4c_1c_2\hat R_1
$$
$$
+ 16c_3\hat H^2 + 4(c_3+c_2^2-4c_1c_3)\hat H + 4c_1^2c_3\,.
$$
As in the case of free motion, equation 
 $H=E$ becomes
$$
p^2_\xi +p^2_\eta +2c_1+ {8c_2\over (\xi +i\eta )^2} + 
{16c_3(\xi -i\eta )\over (\xi +i\eta )^3} -E(4+\xi ^2+\eta ^2)=0,
$$
a superintegrable system in flat space with rearranged 
constants, that separates variables in polar and hyperbolic coordinates.

\begin{itemize}
\item[{\bf [D]}]
$\ds H= 
\frac{\mu^2p^2_\mu-\nu^2p^2_\nu
         +d_1\mu+d_2\nu+d_3(\mu^2+\nu^2)}
     {(\mu+\nu)(2+\mu-\nu)}$.
\end{itemize}
The additional constants of the motion have the form 
\beann
R_1 &=& X_1 + iX_2 - K^2
         - \frac{\mu\nu\Bigl(d_1(\nu-2)+d_2(\mu+2)+2d_3(\nu-\mu+\mu\nu)\Bigr)}
                {(\mu+\nu)(2+\mu-\nu)}\,, \\
R_2 &=& X_1 - iX_2 - \frac{(\mu-\nu)\Bigl((\mu-\nu)(d_1\mu+d_2\nu)
                             -2d_3(\mu^2+\nu^2+\mu\nu(2+\mu-\nu))\Bigr)}
                          {4\mu\nu(\mu+\nu)(2+\mu-\nu)}\,.
\eeann
The corresponding quadratic Poisson algebra can be determined from
\beann
R^2 &=& 4R_1R_2^2
- 4HR_1R_2
+ d_3^2R_1
- 4H^2R_2
+ 2(d_1+d_2)HR_2
- d_1d_2R_2 \\
 & & \ {}
+ 4H^3
- 2(d_1+d_2)H^2
+ \frac14\Bigl((d_1+d_2)^2+d_3(d_2-d_1)\Bigr)H
 - d_3(d_1^2-d_2^2).
\eeann
This classical system also separates in elliptical 
coordinates obtained by choosing new variables defined by the roots of the 
characteristic equation of $R_1+R_2$, that is, the elliptical coordinates
(\ref{coordsD3X1K}) with $b=2i$.
In these variables the Hamiltonian has the form 
$$
H = \frac{p_\omega^2 + p_\varphi^2 + 2(d_1+d_2)(\cos2\varphi-\cosh2\omega)
             + 2(d_1-d_2)(2i\sin2\varphi+\sinh2\omega)
             + 2d_3(\sinh4\omega+2i\sin4\phi)}
         {8(\cos2\varphi-\cosh2\omega)+4(\cosh^22\omega-\cos^22\varphi)}\,.
$$
The corresponding quantum algebra relations are 
\beann
[\hat R,\hat R_1]
 &=& -2\{\hat R_1,\hat R_2\} + 2\hat H\hat R_1 + \hat R_2
      + 2\hat H^2 - \left(d_1+d_2+\sfrac12\right)\hat H
      + \sfrac12d_1d_2 \\{}
[\hat R,\hat R_2]
 &=& 2\hat R_2^2 - 2\hat H\hat R_2 + \sfrac12d_3^2 \\
\hat R^2
 &=& 2\{\hat R_1,\hat R_2^2\}
- 5\hat R_2^2
- 2\hat H\{\hat R_1,\hat R_2\}
+ d_3^2\hat R_1
- 4\hat H^2\hat R_2
+ (2d_1+2d_2+5)\hat H\hat R_2
- d_1d_2\hat R_2 \\
 & & \ {}
+ 4\hat H^3
- (2d_1+2d_2+1)\hat H^2
+ \left(\sfrac14(d_1+d_2)^2+d_3(d_2-d_1)\right)\hat H
- \sfrac14d_3(d_3-d_1^2+d_2^2)
\eeann

As in the case of free motion we observe that equation $H=E$ becomes 
$$
p^2_\xi +p^2_\eta + d_1 + d_2 - 4d_3 
 + \frac{(d_2-d_1)(\xi -i\eta )}
        {\sqrt{(\xi -i\eta )^2+4}}
 + \frac{8d_3(\xi+i\eta)}
   {\sqrt{(\xi -i\eta )^2+4}\Bigl(\xi-i\eta+\sqrt{(\xi-i\eta)^2+4}\Bigr)^2} 
$$
$$
 = (E-d_3)(4+\xi^2+\eta^2)\,,
$$
a superintegrable system in flat space with rearranged 
constants that separates variables in elliptic and hyperbolic coordinates.

\begin{itemize}
\item[{\bf[E]}]
$\ds H = \frac{p^2_\xi+p^2_\eta +c}{4+\xi^2+\eta^2}$.
\end{itemize}
The additional constants of the motion are  
$$
R_1 = X_1 + \frac c4\frac{\eta^2-\xi^2}{4+\xi^2+\eta^2}\,, \qquad
R_2 = X_2 - \frac c2\frac{\xi\eta}{4+\xi^2+\eta^2}\,,
$$
and $K$. The corresponding Poisson algebra relations have the form 
$$
\{K,R_1\} = -R_2,\qquad 
\{K,R_2\} = R_1\,, \qquad 
\{R_1,R_2\} = HK\,,
$$
and the functional relation between these constants is 
$$
R_1^2 + R_2^2 - HK^2 - H^2 + \frac c2H -\frac{c^2}{16}=0\,.
$$

This Hamiltonian separates in all of the four types of separable coordinate 
systems available, and the corresponding expressions for the
Hamiltonian  can be  deduced from [2] by taking $b_3=c$, $b_1=b_2=0$.

The quantum algebra relations are
$$
[\hat K,\hat R_1]=-\hat R_2\,,\qquad 
[\hat K,\hat R_2]=\hat R_1\,,\qquad 
[\hat R_1,\hat R_2]=\hat H\hat K\,,
$$
and the associated operator identity is 
$$
\hat R_1^2+\hat R_2^2 -\hat H\hat K^2 - \hat H^2 + \left(\frac c2+\frac14\right)\hat H
  - \frac{c^2}{16} = 0\,.
$$

\section{Darboux spaces of type four}
\label{sec:Darboux4}

\subsection{The free particle and separating coordinate systems}
\label{subsec:Darboux4free}
With rescaling of the variables $x$ and $y$, the Hamiltonian $H$ can 
be taken in the form 
\be
H_0 = \frac{(e^{x-y} - e^{y-x})^2}
         {e^{x-y}+e^{y-x} +a}\ p_xp_y.
\ee
In coordinates $x=v+iu$, $y=v-iu$, we can write the Hamiltonian as
$$
H_0 = -\frac{\sin^22u(p_u^2+p_v^2)}{2\cos2u+a}\,.
$$
It admits constants of the motion
$$
K =p_v\,\quad
X_1 = e^{2v}\left(-H_0+\cos2u\ p_u^2+\sin2u\ p_up_v\right),\quad 
X_2 = e^{-2v}\left(-H_0+\cos2u\ p_v^2-\sin2u\ p_up_v\right).
$$
These integrals satisfy the polynomial algebra relations 
$$
\{K,X_1\}=2X_1,\qquad 
\{K,X_2\}=-2X_2,\qquad  
\{X_1,X_2\}=-8K^3-4aKH_0.
$$
They are functionally dependent via the relation 
$$
X_1X_2-K^4-aK^2H_0-H_0^2=0.
$$
The corresponding quantum operators are
$$
\hat H_0 = \frac{-\sin^22u}{2\cos2u+a} \ \left(\partial^2_u + \partial^2_v\right)\,,
\qquad \hat X_1 = e^{2v}\left( -\hat H_0 + \cos2u(\partial^2_v + \partial_v) 
                                + \sin2u(\partial_u\partial_v + \partial_u)\right)\,, 
$$
$$
\hat K = \partial_v\,, \qquad 
\hat X_2 = e^{-2v}\left( -\hat H_0 + \cos2u(\partial^2_v - \partial_v) 
                                - \sin2u(\partial_u\partial_v - \partial_u)\right)\,.
$$
Their algebra is determined by the relations 
$$
[\hat K,\hat X_1] = 2\hat X_1,\qquad 
[\hat K,\hat X_2] = -2\hat X_2,\qquad  
[\hat X_1,\hat X_2]= -8\hat K^3-4a\hat K\hat H_0-4\hat K,
$$
and the  operator identity is 
$$
\frac12\{\hat X_1,\hat X_2\}
-\hat K^4
-a\hat H_0\hat K^2
-5\hat K^2
-\hat H_0^2
-a\hat H_0
=0\,.
$$

The line element
$ds^2= (2\cos u+a) (du^2+dv^2)/\sin ^22u$ can be realized as a
two-dimensional surface embedded in $E(2,1)$ by
 (assuming $a>2$)
$$
X=\sqrt{a+2\cos 2u}v,\qquad 
Y-T=\sqrt{a+2\cos 2u},
$$
$$
Y+T= {(a-2)\over \sqrt{2(a+2)}} 
[\Pi (\chi ,\sqrt{{a-2\over a+2}}{2\over (r_1+1)} 
,p)+\Pi (\chi ,\sqrt{{a-2\over a+2}}{2\over (r_2+1)}
,p)]-\sqrt{a+2\cos 2u}v^2,
$$
where  
$$
\sin\chi =\sqrt{{(a+2)(\cos 2u+1)\over 2(a+2\cos 2u)}},\qquad
p= {2\over \sqrt{a+2}},
$$
and $\Pi $ is an elliptic integral of the third kind \cite{EMOT1}. 
Then $ds^2=dX^2+dY^2-dT^2$.

Just as we have done in other cases, we wish to determine all the essentially 
different separable coordinate systems for the free classical or quantum 
particle. To do this we need to consider a general quadratic constant of the 
form 
$
\lambda =aX_1+bX_2+cK^2$.
Under the adjoint action of $\exp (\alpha K)$,   $X_1$ and $X_2$ transform 
according to  
$$
X_1\rightarrow \exp (-2\alpha )X_1,\qquad  
X_2\rightarrow \exp (2\alpha )X_2.
$$
If we regard two such quadratic expressions as equivalent if they are
related by a combination of group motions and the discrete transformation 
observed above, then the equivalence classes of these expressions  can be chosen
to have the following representatives:
\be
K^2\,,\qquad
X_2\,,\qquad
\gamma X_2+K^2\,,\qquad
X_1+X_2+\gamma K^2.
\label{freesep4}
\ee
In the last of these are three cases to distinguish:  $\gamma=0$, 
$\gamma=2$ and $\gamma\neq 0,2$. 
The various separable systems involved can now be
computed.

\subsubsection{Coordinates associated with $K^2$}
These are the coordinates associated with the ignorable coordinate $v$ and the
Hamiltonian has already been given in the $u,v$ coordinates. The Hamilton-Jacobi equation is
$$
-\frac{\sin^22u}{2\cos2 u+a}\left(\pdsq S v + \pdsq S u\right)
      = E\,.
$$
It has typical solutions 
\beann
S(u,v)
 &=& -i\log\left(i(c^2\cos2 u-E)+c\sqrt{E(a+2\cos2 u)+c^2\sin^22 u}\right) \\
 & & \ {}
     + \frac1{2c}\sqrt{E(a+2)} \ \mbox{arctanh}\left(
\frac{E(a+1)+c^2+(E-c^2)\cos2 u}{\sqrt{E(a+2)(E(a+2\cos2 u)+c^2\sin^22 u})}\right) \\
 & & \ {}
     + \frac1{2c}\sqrt{E(a-2)} \ \mbox{arctanh}\left(
\frac{E(a-1)+c^2+(E+c^2)\cos2 u}{\sqrt{E(a-2)(E(a+2\cos2 u)+c^2\sin^22 u})}\right)
     + c v\,.
\eeann
The corresponding Schr\"odinger equation is
$$
\frac{\sin^22 u}{2\cos2 u+a}\left(\pdtwo\Psi v + \pdtwo\Psi u\right)
      = E\Psi\,,
$$
which has the solution
$$
\Psi = {}_2F_1\left(\frac12(\lambda-\epsilon_+-\epsilon_-),
                    \frac12(\lambda+\epsilon_++\epsilon_-),
                    \epsilon_++\frac12,\sin^2 u\right)e^{\lambda v}\,,
$$
where
\be
\label{epsilonpm}
\epsilon_\pm=\frac12+\frac12\sqrt{1-(a\pm2)E}\,,
\ee
%
and  ${}_2F_{1}$ is a Gaussian hypergeometric function \cite{EMOT}.

\subsubsection{Coordinates associated with $X_2$}

If we choose new coordinates 
\be
x=\log\left(\frac12(\mu -i\nu )\right)\,, \qquad y=\log\left(\frac12(\mu +i\nu )\right)\,,
\ee
then the
Hamiltonian takes the rational form 
$$
H_0 = - \frac{4\mu ^2\nu ^2(p^2_\mu +p^2_\nu )}
           {(a+2)\mu ^2+(a-2)\nu ^2}\,.
$$

In this case the corresponding choice of coordinates has 
already been given, and the  quadratic constant in these coordinates is  
$$
X_2= {4(a+2)\mu ^2p^2_\mu -4(a-2)\nu ^2p^2_\nu \over (a+2)\mu ^2+(a-2)\nu ^2}.
$$
The Hamilton-Jacobi equation 
$$
 - \frac{\ds4\mu ^2\nu ^2\left(\pdsq S\mu  + \pdsq S\nu \right)}{(a+2)\mu ^2+(a-2)\nu ^2} = E
$$
has solution
\beann
S(\mu ,\nu ) &=& i\sqrt{(a-2)E+\lambda \mu ^2}
           - i\sqrt{(a-2)E} \ \mbox{arctanh}\sqrt{\frac{(a-2)E+\lambda \mu ^2}{(a-2)E}} \\
       & & \quad {} + \sqrt{(a+2)E-\lambda \nu ^2}
           - \sqrt{(a+2)E} \ \mbox{arctanh}\sqrt{\frac{(a+2)E-\lambda \nu ^2}{(a+2)E}}
\eeann
The corresponding Schr\"odinger equation  has
Bessel function solutions of the form
$$
\Psi=\sqrt{\mu \nu } \ C_{\frac12\sqrt{1-E(a-2)}}\left(\sfrac12\sqrt{\lambda} \ \mu \right)
\ C_{\frac12\sqrt{1-E(a+2)}}\left(\sfrac12\sqrt{\lambda} \ i\nu \right)\,.
$$
\subsubsection{Coordinates associated with $\gamma X_2+K^2$}
In the case
of the third representative the transformation  
\be
\label{coordsD4X2Ksq}
\mu=c\cosh\omega \cos\varphi\,,\qquad
\nu=c\sinh\omega \sin\varphi
\ee
gives the classical Hamiltonian  
$$
H= \frac{4(p^2_\omega +p^2_\varphi )}
        {(a-2)(\sech^2\omega-\sec^2\varphi)
           - (a+2)(\cosech^2\omega + \cosec^2\varphi)}\,.
$$
The classical constant associated with this coordinate system is 
\beann
\lefteqn{-{c^2\over 4}X_2+K^2} \\
 &=& \frac{\left((a-2)\sec^2\varphi+(a+2)\cosec^2\varphi\right)p_\omega^2
      + \left((a-2)\sech^2\omega-(a+2)\cosech^2\omega\right)p_\varphi^2}
        {(a-2)(\sech^2\omega-\sec^2\varphi)
           - (a+2)(\cosech^2\omega + \cosec^2\varphi)}\,.   
\eeann
The Hamilton-Jacobi equation in these coordinates is
$$
\frac{4(p^2_\omega +p^2_\varphi )}
        {(a-2)(\sech^2\omega-\sec^2\varphi)
           - (a+2)(\cosech^2\omega + \cosec^2\varphi)} = E
$$
and has solutions of the form
\beann
S(\omega,\varphi)
 &=& \frac14\sqrt\lambda\log\left(\sqrt\lambda \ (\lambda\cos2\varphi+2aE)
                    +\lambda\sqrt{8E+4aE\cos2\varphi-\lambda\sin^22\varphi}\right) \\
 & & \ {}
     -\frac14\sqrt{(a+2)E} \ \mbox{arctanh}\left(
       \frac{2Ea+8E-\lambda+\cos2\varphi(\lambda+2Ea)}
                   {2\sqrt{(a+2)E(8E+4aE\cos2\varphi-\lambda\sin^22\varphi)}}\right) \\  
 & & \ {}
     -\frac14\sqrt{(a-2)E} \ \mbox{arctanh}\left(
       \frac{-2Ea+8E-\lambda+\cos2\varphi(-\lambda+2Ea)}
                   {2\sqrt{(a-2)E(8E+4aE\cos2\varphi-\lambda\sin^22\varphi)}}\right) \\
 & & \ {}
     +\frac14\sqrt\lambda\log\left(\sqrt\lambda \ (\lambda\cosh2\omega-4E)
                     +\lambda\sqrt{\lambda\sinh^22\omega-8E\cosh2\omega-4aE}\right) \\  
 & & \ {}
     +\sqrt{a+2} \ \mbox{arctan}\left(
       \frac{4E+\lambda+4aE+\cosh2\omega(4E-\lambda)}
                   {2\sqrt{(a+2)E(\lambda\sinh^2\omega-8E\cosh2\omega-4aE)}}\right) \\
 & & \ {}
     +\sqrt{a-2} \ \mbox{arctan}\left(
       \frac{-4E+\lambda+4aE+\cosh2\omega(4E+\lambda)}
                   {2\sqrt{(a-2)E(\lambda\sinh^2\omega-8E\cosh2\omega-4aE)}}\right)\,.
\eeann
The Schr\"odinger equation has the form
$$
\frac{\ds4\left(\pdtwo\Psi\omega+\pdtwo\Psi\varphi\right)}
     {(a-2)\left(\sech^2\omega-\sec^2\varphi\right)
          - (a+2)\left(\cosech^2\omega+\cosec^2\varphi\right)}
   = E\Psi
$$
with corresponding solutions
$$
\Psi = \left(\sin\varphi\sinh\omega\right)^{\epsilon_-}
         \left(\cos\varphi\cosh\omega\right)^{\epsilon_+}
       {}_2F_1\left(\frac{\epsilon_++\epsilon_--\lambda}2,
                    \frac{\epsilon_++\epsilon_-+\lambda}2,
                    \epsilon_-+\frac12,\sin^2\varphi\right)
$$
$$
\times \       {}_2F_1\left(\frac{\epsilon_++\epsilon_--\lambda}2,
                    \frac{\epsilon_++\epsilon_-+\lambda}2,
                    \epsilon_-+\frac12,-\sinh^2\omega\right)
$$
with $\epsilon_\pm$ defined by (\ref{epsilonpm}).

\subsubsection{Coordinates associated with $X_1+X_2+\gamma K^2$}

For the coordinates corresponding to the fourth representative we make the transformation 
$u =\arctan(\exp\alpha )$, $v =\beta/2$, so our Hamiltonian has the 
form
\be
\label{H3alphabeta}
H = -4 \frac{p^2_\alpha + \sech^2\alpha p^2_\beta}
            {a-2\tanh\alpha}\,.
\ee
This can be realised in terms of projective coordinates on a two-dimensional 
complex sphere via
$s_1=\cosh\alpha \cosh\beta$, $s_2=i\cosh\alpha \sinh\beta$,
$s_3=i\sinh\alpha$
where $s_1^2+s_2^2+s_3^2=1$. The Hamiltonian can  be written as
$$
H = 4\frac{J_1^2+J_2^2+J_3^2}
          {\frac{2is_3}{\sqrt{s^2_1+s^2_2}} + a}\,.
$$
These two ways of realising the classical Hamiltonian 
are useful in determining 
the various possible separable coordinate systems. 

We consider the most general case 
first, i.e., $\gamma \neq 0,2$. 
We make use of the transformation equations 
$$
\sinh\alpha  = i\frac{XY+1}{2\sqrt{XY}}\,,
$$
$$
\tanh\beta  = 
\frac{2\sqrt{(A_+X+A_-)(A_-X+A_+)(A_+Y+A_-)(A_-Y+A_+)}}
     {(A^2_++A^2_-)(XY+1) + 2A_+A_-(X+Y)}\,,
$$
applied to (\ref{H3alphabeta}) to give 
classical Hamiltonian in the form
$$
H = - 16XY\frac{X(A_+X+A_-)(A_-X+A_+)p^2_X-Y(A_+Y+A_-)(A_-Y+A_+)p^2_Y}
               {A_+A_-(X-Y)\Bigl((a+2)XY-a+2\Bigr)}\,.
$$
The corresponding classical constant associated with this coordinate system is 
\beann
\lefteqn{X_1+X_2+2\frac{A^2_++A^2_-}{A^2_+-A^2_-} K^2} \\
 &=& 16\frac{(A_+X+A_-)(A_-X+A_+)\Bigl(a(A_+Y+A_-)(A_-Y+A_+)-2A_+A_-(Y^2-1)\Bigr)X^2p_X^2}
          {A_+A_-(A_+^2-A_-^2)(X-Y)\Bigl((a+2)XY-a+2\Bigr)} \\
 & & \ {}
 - 16\frac{(A_+Y+A_-)(A_-Y+A_+)\Bigl(a(A_+X+A_-)(A_-X+A_+)-2A_+A_-(X^2-1)\Bigr)Y^2p_Y^2}
          {A_+A_-(A_+^2-A_-^2)(X-Y)\Bigl((a+2)XY-a+2\Bigr)}\,.
\eeann
The Hamilton-Jacobi equation has the form
$$
- 16XY\frac{X(A_+X+A_-)(A_-X+A_+)\pdsq SX+Y(A_+Y+A_-)(A_-Y+A_+)\pdsq SY}
               {A_+A_-(X-Y)\Bigl((a+2)XY-a+2\Bigr)}
$$
and solutions
$$
S(X,Y) = \frac1{\sqrt{A_+A_-}}\left(
            \lambda_X\int\frac1X\sqrt{\frac{a_X-X}{(b-X)(c-X)}} \ dX
            + \lambda_Y\int\frac1Y\sqrt{\frac{a_Y-Y}{(b-Y)(c-Y)}} \ dY
                              \right)
$$
where $\lambda_X-\lambda_Y=-(a+2)EA_+A_-/16$, $a_X=(a-2)E\lambda_X/16$,
$a_Y=(a-2)E\lambda_Y/16$, $b=-A_+/A_-$, $c=-A_-/A_+$.

A further change of coordinates
$$
X = -\frac1k\mbox{sn}^2(\alpha'+iK',k)\,,\qquad
Y = -\frac1k\mbox{sn}^2(\beta'+iK',k)\,, \qquad k = \frac{A_+}{A_-}
$$
is convenient for writing the Schr\"odinger equation
$$
\frac{\ds16\left(\pdtwo\Psi{\alpha'}+\pdtwo\Psi{\beta'}\right)}
     {(a+2)k^4\left(\mbox{sn}^2(\alpha',k)-\mbox{sn}^2(\beta',k)\right)+k^2(a-2)}
   = E\Psi\,.
$$
The separated equations are versions of Lame's equation \cite{WW}.  Indeed if we look
for solutions of the form $\Psi=A(\alpha')B(\beta')$ then
\beann
\pdtwo{A(\alpha')}{\alpha'}
  + \left(-\frac1{16}k^4E(a+2)\mbox{sn}^2(\alpha',k)-\lambda_1\right)A(\alpha') = 0\,, \\
\pdtwo{B(\beta')}{\beta'}
  + \left(-\frac1{16}k^4E(a+2)\mbox{sn}^2(\beta',k)-\lambda_2\right)B(\beta') = 0
\eeann
where $\lambda_1-\lambda_2=-E(a-2)k^2/16$.  Solutions of these separation
equations can be represented as Riemann $P$ functions \cite{MagO} of the form
$$
P(z) = \left(
   \begin{array}{ccccc}
      0 & 0 & k^{-2} & \infty & \\
      0 & 0 & 0 &
          \frac14\left(1-\frac12\sqrt{4+k^2E(a+2)}\right) & \mbox{sn}^2(z,k) \\
      \frac12 & \frac12 & \frac12 &
          \frac14\left(1+\frac12\sqrt{4+k^2E(a+2)}\right) &
   \end{array}
       \right)
$$
for $z=\alpha',\beta'$.

The case $\gamma =0$ can easily be deduced by putting $A_+=iA_-$, as can be seen 
from the expression for the associated classical constant.

If $\gamma=2$ then a convenient choice of coordinates is 
\be
x=\log \Bigl(\tan (\varphi'-i\omega')\Bigr)\,, \qquad
y=\log \Bigl(\tan (\varphi'+i\omega')\Bigr)\,.
\ee
The corresponding classical Hamiltonian  has the form
$$
H = -\frac{p_{\varphi'}^2+p_{\omega'}^2}
         {\ds\frac{a+2}{\sinh^22\omega'}+\frac{a-2}{\sin^22\varphi'}}\,.
$$
The classical constant is
$$
X_1+X_2+2K^2
 = aH + \frac{(a+2)\sin^22\varphi'p_{\varphi'}^2-(a-2)\sinh^22\omega'p_{\omega'}^2}
             {(a+2)\sin^22\varphi'+(a-2)\sinh^22\omega'}
$$
The Hamilton-Jacobi equation in these coordinates is
$$
-\frac{\pdsq S{\varphi'} + \pdsq S{\omega'}}
     {\ds\frac{a+2}{\sinh^22\omega'}+\frac{a-2}{\sin^22\varphi'}} = E
$$
which has solutions
\begin{eqnarray*}
S(\varphi',\omega')
 &=& \frac i2\sqrt\lambda \
      \mbox{arctan}\sqrt{\frac{(a-2)E}\lambda \ \sec^22\varphi' + \tan^22\varphi'} \\
 & & \quad {} - \frac i2\sqrt{(a-2)E} \ 
      \mbox{arctanh}\sqrt{\sec^22\varphi' + \frac\lambda{(a-2)E}\tan^22\varphi'} \\
 & & \quad {} + \frac i2\sqrt\lambda \ 
      \mbox{arctan}\sqrt{\frac{(a+2)E}\lambda \ \sech^22\omega' - \tanh^22\omega'} \\
 & & \quad {} - \frac i2\sqrt{(a+2)E} \ 
      \mbox{arctanh}\sqrt{\sech^22\omega' - \frac\lambda{(a+2)E}\tanh^22\omega'}
\end{eqnarray*}
The corresponding Schr\"odinger equation is
$$
-\frac{\ds\pdtwo\Psi{\varphi'}+\pdtwo\Psi{\omega'}}
     {\ds\frac{a+2}{\sinh^22\omega'}+\frac{a-2}{\sin^22\varphi'}}
  = E\Psi\,,
$$
which has solutions of the form
$$
\Psi = \sqrt{\sin2\varphi'\sinh2\omega'} \ P_\nu^{\frac12\sqrt{1-(a-2)E}}(\cos2\varphi')
     P_\nu^{\frac12\sqrt{1-(a+2)E}}(\cosh2\omega')
$$
where $P_\nu^\mu(z)$ is a solution of Legendre's equation.

This  completes the list of possible coordinate systems which are 
inequivalent and separable for this particular Hamiltonian. We notice in 
particular that the equation $H-E=0$ can be written in the equivalent forms
$$
\mu^2(p_\mu^2+p_\nu^2) + \frac14E\left(a-2+(a+2)\frac{\mu^2}{\nu^2}\right) = 0,\qquad 
J_1^2+J_2^2+J_3^2-E\left(\frac{2is_3}{\sqrt{s^2_1+s^2_2}} + a\right)=0\,,
$$
both superintegrable systems on the complex two-sphere, the
first of which is written in horospherical coordinates.

\subsection{Superintegrability for Darboux spaces of type four.}
\label{subsec:superint4}
There are various possibilities for the potential in this case: [{\bf A}],  [{\bf B}],  [{\bf C}],  [{\bf D}].

\begin{itemize}
\item[{\bf [A]}]
$\ds H = -\frac{4\mu ^2\nu ^2}{(a+2)\mu ^2+(a-2)\nu ^2}
     \left(p^2_\mu +p^2_\nu +a_1+a_2\left(\frac1{\mu ^2}+\frac1{\nu ^2}\right) + a_3(\mu ^2+\nu ^2)\right)$.
\end{itemize}
The additional constants of the motion have the form 
\beann
R_1 &=& K^2 + a_1(\mu ^2+\nu ^2) + a_3(\mu ^2+\nu ^2)^2\,, \\
R_2 &=& X_2 + \frac{2a_1\Bigl((a+2)\mu ^2-(a-2)\nu ^2\Bigr)+16a_2+4a_3\Bigl((a+2)\mu ^4-(a-2)\nu ^4\Bigr)}
                   {(a+2)\mu ^2+(a-2)\nu ^2}\,.
\eeann
The corresponding quadratic algebra relations are determined by 
\beann
R^2 &=& 
16R_1R_2^2
-256a_3R_1^2
-64a_1R_1R_2
-256aa_3HR_1
-1024a_2a_3R_1
+64a_1HR_2 \\
& & \ {}
-256a_3H^2
-64a_1(a+2)H
-256a_1^2a_2\,.
\eeann
This Hamiltonian admits a separation of variables in coordinates corresponding 
to the equivalence first, second and third classes of 
Section \ref{subsec:Darboux4free}. For the second this is covered by the 
choice of coordinates $\mu ,\nu $.
\begin{itemize}
\item[(i)] 
For coordinates corresponding to the first equivalence class, we  obtain the Hamiltonian in the form
$$
H = - \frac{\sin^22u\left(p_u^2+p_v^2+4a_1e^{2v} + 4a_2\cosec^22u
             + 4a_3e^{4v}\right)}{2\cos2u+a} \,.
$$
\item[(ii)]
For coordinates corresponding to the third representative (\ref{coordsD4X2Ksq})
the Hamiltonian takes form 
\beann
H &=& \frac{4(p_\omega^2+p_\varphi^2) + 4a_1c^2(\cosh^2\omega-\cos^2\varphi)}
         {(a-2)\left(\sech^2\omega-\sec^2\varphi\right)
             -(a+2)\left(\cosech^2\omega+\cosec^2\varphi\right)} \\
& & {} \ + \frac{16a_2\left(\cosech^22\omega+\cosec^22\varphi\right)
             + a_3c^4(\sinh^22\omega+\sin^22\varphi)}
         {(a-2)\left(\sech^2\omega-\sec^2\varphi\right)
             -(a+2)\left(\cosech^2\omega+\cosec^2\varphi\right)}
\eeann
\end{itemize}
The quantum algebra relations are 
\beann
[\hat R,\hat R_1]
  &=& -8\{\hat R_1,\hat R_2\} - 16\hat R_2 - 32a_1\hat H\,, \\
{}[\hat R,\hat R_2]
  &=& 8\hat R_2^2 - 256a_3\hat R_1 - 128aa_3\hat H
       - 32(a_1^2+4a_3+16a_2a_3)
\eeann
together with the operator relation 
\beann
\hat R^2 &=&
8\{\hat R_1,\hat R_2^2\}
-256a_3\hat R_1^2
-80\hat R_2^2
-256aa_3\hat H\hat R_1
-64(16a_2a_3+a_1^2+4a_3)\hat R_1 \\
 & & \ {}
+64a_1\hat H\hat R_2
-256a_3\hat H^2
+64a(4a_3-a_1^2)\hat H
+128(a1^2+4a_3+8a_2a_3-2a_1^2a_2).
\eeann

As in the case of free motion we observe that  the equation  $H=E$ is
$$
p_\mu^2+p_\nu^2 + a_1 + \frac{a_2-\frac14(a-2)E}{\mu^2} + \frac{a_2-\frac14(a+2)E}{\nu^2}
     + a_3(\mu^2+\nu^2) = 0\,,
$$
a superintegrable system in flat space with rearranged 
constants, that separates  in elliptic and hyperbolic coordinates.

\begin{itemize}
\item[{\bf [B]}]
$\ds H = -\frac{\ds\sin^22u \left(p^2_ v +p^2_u 
            + \frac{b_2}{\sinh^2 v} + \frac{b_3}{\cosh^2 v}\right)+b_1}
               {2\cos2u +a}$
\end{itemize}
The additional constants are
\beann
R_1 &=& X_1 + X_2 \\
 & & \ {} + 
  \frac{\ds 2b_1\cosh2 v + (b_2+b_3)(4-a^2)
           + (\cos4u +2a\cos2u +3)
                \left(\frac{b_2}{\sinh^2 v}-\frac{b_3}{\cosh^2 v}\right)}
       {2\cos2u +a} \\
R_2 &=& K^2+ \frac{b_2}{\sinh^2 v} + \frac{b_3}{\cosh^2 v}\,.
\eeann
The quadratic algebra is given by 
\beann
R^2 &=&
16R_1^2R_2
-64R_2^3
-64aHR_2^2
+64(2b_3-2b_2-b_1)R_2^2
+32a(b_2+b_3)R_1R_2 \\
 & & \ {}
-64H^2R_2
+64(b_2+b_3)HR_1
+128a(b_3-b_2)HR_2 \\
 & & \ {}
-16\Bigl((4-a^2)(b_2+b_3)^2+8b_1(b_2-b_3)\Bigr)R_2
+128(b_3-b_2)H^2
-64b_1(b_2+b_3)^2\,.
\eeann

This Hamiltonian admits a separation of variables in coordinate systems 
corresponding to the first and fourth equivalence classes of (\ref{freesep4}). 
The defining
expressions have already been given in terms of coordinates for the
first.  For the fourth, we distinguish  two  cases.
\begin{itemize}
\item[(i)]
$\gamma\neq2$
\beann
H &=& 16XY \ \frac{X(A_-X-A_+)(A_+X-A_-)p^2_X+Y(A_-Y+A_+)(A_+Y+A_-)p^2_Y}
                  {A_+A_-(X-Y)(a-2-(a+2)XY)} \\
  & & {} \
 + \frac{\ds b_1(XY+1) + \frac{4b_2(A^2_--A^2_+)XY}{(A_+Y+A_-)(A_+X+A_-)}
                  +\frac{4b_3(A^2_--A^2_+)XY}{(A_-Y+A_+)(A_-X+A_+)}}
        {a-2-(a+2)XY} \\
\eeann
\item[(ii)]
$\gamma=2$
$$
H = -\frac{\ds p_{\varphi'}^2+p_{\omega'}^2
              + b_1\left(\frac1{\sinh^22\omega'}+\frac1{\sin^22\varphi'}\right)
              + \frac{4b_2}{\cos^22\varphi'} + \frac{4b_3}{\cosh^22\omega'}}
         {\ds \frac{a+2}{\sinh^22\omega'} + \frac{a-2}{\sin^22\varphi'}}
$$
\end{itemize}
The corresponding quantum algebra relations are 
\beann
[\hat R,\hat R_1]
  &=& -8\hat R^2_1 + 96\hat R^2_2 + 64a\hat H\hat R_2
       - 16a(b_2+b_3)\hat R_1
 + 64(2b_2-2b_3+b_1+3)\hat R_2 \\ 
  & & {} \
 + 32\hat H^2
 + 32a(2b_2-2b_3+1)\hat H \\
  & & {} \
 + 64b_1(b_2-b_3) - 8(a^2-4)(b_2+b_3)^2
 + 32(b_1+2b_2-2b_3)\,, \\
{}[\hat R,\hat R_2]
  &=& 8\{\hat R_1,\hat R_2\} + 16a(b_2+b_3)\hat R_2 - 16\hat R_1
         + 32(b_2+b_3)\hat H - 16a(b_2+b_3)\,, \\
R^2 &=& 
- 64 \hat R_2^3
+ 8 \{\hat R_1^2,\hat R_2\}
- 64a \hat H\hat R_2^2
- 64 \hat H^2\hat R_2
- 80 \hat R_1^2
- 64(2b_2-2b_3+b_1+7)\hat R_2^2 \\
 & & {} \ 
+ 16a(b_2+b_3) \{\hat R_1,\hat R_2\}
+ 64(b_2+b_2) \hat H\hat R_1
+ 64a(2b_3-2b_2-1) \hat H\hat R_2 \\
 & & {} \
- 160a(b_2+b_3) \hat R_1
+ 16\Bigl((a^2-4)(b_2+b_3)^2 + 8(b_1+1)(b_3-b_2) - 4b_1 + 32\Bigr)\hat R_2 \\
 & & {} \
+ 128(b_3-b_2+1) \hat H^2
+ 128a(b_2-b_3+1)\hat H \\
 & & {} \
+ (b_2+b_3)^2(128-80a^2-64b_1) - 128(b_1+2)(b_3-b_2-1) - 256.
\eeann
As in the case of free motion, equation $H-E=0$ is
$$
J_1^2 + J_2^2 + J_3^2  + \frac{2b_1}{\sqrt{s^2_1+s^2_2}\left(s_1+\sqrt{s^2_1+s^2_2}\right)} 
   + \frac{2b_2}{\sqrt{s^2_1+s^2_2}\left(s_1-\sqrt{s^2_1+s^2_2}\right)} + b_3
$$ 
$$
-E\left(\frac{2is_3}{\sqrt{s^2_1+s^2_2}}+a\right)=0\,,
$$
a superintegrable system on the complex sphere that
separates variables in spherical, elliptic and degenerate elliptic type 1 
coordinates.

\begin{itemize}
\item[{\bf [C]}]
$\ds H = - \frac{\ds p_{\varphi'}^2+p_{\omega'}^2+\frac{c_1}{\cos^2\varphi'}
                  +\frac{c_2}{\cosh^2\omega'}
                  +c_3\left(\frac1{\sin^2\varphi'}-\frac1{\sinh^2\omega'}\right)}
               {\ds \frac{a+2}{\sinh^22\omega'}+\frac{a-2}{\sin^22\varphi'}}\,.
$
\end{itemize}
These are coordinates associated with $\gamma=2$ in the fourth representative
from (\ref{freesep4}).
The constants of the motion associated with this Hamiltonian are
\beann
R_1 &=& X_1 + X_2 + 2K^2 + aH \\
 & & \ {} + \frac{\ds \frac{a+2}{\sinh^22\omega'}
          \left(\frac{c_3}{\sin^2\varphi'}+\frac{c_1}{\cos^2\varphi'}\right)
                   + \frac{a-2}{\sin^22\varphi'}
          \left(\frac{c_3}{\sinh^2\omega'}-\frac{c_2}{\cosh^2\omega'}\right)}
                 {\ds \frac{a+2}{\sinh^22\omega'}+\frac{a-2}{\sin^2\omega'}}\,. \\
R_2 &=& X_1 - X_2 + \frac1{\ds\frac{a+2}{\sinh^22\omega'}+\frac{a-2}{\sin^2\omega'}} \\ 
 & & \ {} \times \left[ \frac{a+2}{\sinh^22\omega'}
          \left(c_1\cosh2\omega'\tan^2\varphi'-c_2\cos2\varphi'
 -\frac{c_3\Bigl(2\cos^2\varphi'(\sinh^2\omega'-\sin^2\varphi')\Bigr)+1}{\sin^2\varphi'}\right)
 \right. \\
 & & \left.  {}   + \frac{a-2}{\sin^22\varphi'}
          \left(c_2\cos2\varphi'\tanh^2\omega' + c_1\cosh2\omega' 
 -\frac{c_3\Bigl(2\cosh^2\omega'(\sinh^2\omega'-\sin^2\varphi)+1\Bigr)}{\sinh^2\omega'}\right)
                 \right]\,.
\eeann
They satisfy the quadratic algebra determined by the identity
\beann
R^2 &=& 
16R_1^3
- 16R_1R_2^2
- 32aHR_1^2
+ 32(c_2-c_1)R_1^2
+ 16(a^2-4)H^2R_1 \\
 & & \ {}
+ 32((a+2)c_1-(a-2)c_2+4c_3)HR_1 \\
 & & \ {}
- 16(2c_3^2-c_1^2-c_2^2+6c_3(c_1+c_2)+4c_1c_2)R_1
- 32(c_2-c_3)(c_1-c_3)R_2 \\
 & & \ {}
- 16((a+2)(c_1-c_3)^2+(a-2)(c_2-c_3)^2)H 
- 32(c_1-c_2)(3c_3^2-c_1c_2-c_3(c_1+c_2))
\eeann

The Hamiltonian admits a separation of variables in a number of coordinates 
systems corresponding to various combinations of the operators $R_1$ and 
$R_2$.  We exhibit the various possibilities.

\begin{itemize}
\item[(i)]
For the constant $R_1-R_2$, the associated separable coordinates are those
corresponding to the third representative in (\ref{freesep4}) with $\gamma=1$.  
In these coordinates, the Hamiltonian is
$$
H = \frac{\ds4(p^2_\omega +p^2_\varphi)
            +\frac{c_1+c_2+2c_3}{2\sinh^22\omega}
            -\frac{(c_1+c_2)\cosh2\omega}{2\sinh^22\omega}
            +\frac{c_3\cos2\varphi}{\sin^22\varphi}}
         {\ds (a-2)\left(\frac1{\cosh^2\omega}-\frac1{\cos^2\varphi}\right)
            -(a+2)\left(\frac1{\sinh^2\omega}+\frac1{\sin^2\varphi}\right)}\,.
$$
\item[(ii)]
In coordinates corresponding to rotations of the fourth representative
in (\ref{freesep4}) with $\gamma \neq 0,2$, that is,  
$B^2_+X_1+(B^2_+-B^2_-)X_2+(2B^2_+-B^2_-)K^2$,
the corresponding Hamiltonian has the form 
$$
H = 16\left[
   - X(B_\mp+X)(B_\pm+X)p_X^2+Y(B_\mp+Y)(B_\pm+Y)p_Y^2
   + \frac{c_1}4\left(\frac1Y-\frac1X\right)
\right.
$$
$$
 \left. \left. {}
   +\frac{c_2}4(X-Y)+\frac{c_3}4(B_\mp^2-B_\pm^2)\left(\frac1{1+B_\mp Y}-\frac1{1+B_\mp X}
                   +\frac1{1+B_\pm Y}-\frac1{1+B_\pm X}\right)\right]\right/
$$
$$
 \left[(B_\mp^2-B_\pm^2)\left(\frac{a-2}{1+B_\pm X}-\frac{a-2}{1+B_\pm Y}
                                     +\frac{a+2}{1+B_\mp Y}-\frac{a+2}{1+B_\mp X}\right)
 \right.
$$
$$
 \left. {} + \left(\frac{a-2}X-\frac{a-2}Y+(a+2)(X-Y)\right)\right]\,.
$$

Here, $B_\pm=B_+/B_-$ and $B_\mp=B_-/B_+$.
The Hamiltonian associated with $R_2$ can be obtained
from this last case by taking $B_-=\sqrt{2}
B_+$.
\end{itemize}

The  quantum algebra relations are 
\beann
[\hat R,\hat R_1]
 &=& 8\{\hat R_1,\hat R_2\} + 16\hat R_2 + 16(c_1-c_3)(c_2-c_3)\,, \\
{}[\hat R,\hat R_2]
 &=& 24\hat R_1^2 - 8\hat R_2^2 - 32a\hat H\hat R_1
      + 8(a^2-4)\hat H^2 + 32(c_1-c_2-\sfrac32)\hat R_1 \\
 & & {} \ + 16((a+2)c_1-(a-2)c_2+a+64c_3)\hat H \\
 & & {} \ + 8c_1^2 + 8c_2^2 - 16c_3^2 - 32c_1c_2 - 48c_3(c_1+c_2) + 16(c_1-c_2)\,.
\eeann
The operator identity is
\beann
\hat R^2 &=&
  16 \hat R_1^3
- 8 \{\hat R_1,\hat R_2^2\}
+ 32(c_2-c_1-\sfrac72)\hat R_1^2
- 80 \hat R_2^2
+ 16(a^2-4) \hat H^2\hat R_1 \\
 & & {} \ 
+ 32\Bigl((a+2)c_1-(a-2)c_2+4c_3+a\Bigr)\hat H\hat R_1
+ 16\Bigl(c_1^2+c_2^2-2c_3^2-6c_3(c_1+c_2) \\
 & & {} \ -4c_1c_2+2(c_1-c_2)-8\Bigr)\hat R_1
- 32(c_2-c_3)(c_1-c_3)\hat R_2
+ 16(a^2-4) \hat H^2 \\
 & & {} \
- 16\Bigl((a+2)((c_1-c_3)^2-2c_1)+(a-2)((c_2-c_3)^2+2c_2)-8c_3-4a\Bigr)\hat H \\
 & & {} \
- 32(c_1-c_2)(3c_3^2-c_1c_2-c_3(c_1+c_2)) \\
 & & {} \
+ 32(c_1^2+c_2^2-4c_3(c_1+c_2)-2c_1c_2+2c_1-2c_2).
\eeann

As in the case of free motion, the equation $H=E$ is
$$
J_1^2 + J_2^2 + J_3^2
  - \frac{i(c_1+c_2+2c_3)s_1}{4\sqrt{s_2^2+s_3^2}}
  + \frac{i(c_1-c_2)(s_1+is_2-s_3)}{4\sqrt2\sqrt{(s_1+is_2)(s_3-is_2)}}
$$
$$
  {}+ \frac{(2c_3-c_1-c_2)(s_1+is_2+s_3)}{\sqrt{(s_1+is_2)(s_3+is_2)}}
  + \frac{i(c_1-c_2)}{4\sqrt2}
  - E\left(a+\frac{2is_1}{\sqrt{s_2^2+s_3^2}}\right) = 0\,.
$$
which is  a superintegrable system on the complex sphere, with 
rearranged constants, that separates variables in elliptic and degenerate 
elliptic coordinates of type 1.

\begin{itemize}
\item[{\bf [D]}]
$\ds H = -\frac{\ds4\mu^2\nu^2\left[p^2_\mu+p^2_\nu+d\left(\frac1{\mu^2}+\frac1{\nu^2}\right)\right]}
               {(a+2)\mu^2+(a-2)\nu^2}$.
\end{itemize}
This Hamiltonian admits three classical constants of the motion  
$$
R_1 = X_1 + \frac{d(\mu^2 +\nu^2)^2}{(a+2)\mu^2+(a-2)\nu^2}\,,\quad
R_2 = X_2 + \frac{16d}{(a+2)\mu^2+(a-2)\nu^2}\,,\quad
K = \mu p_\mu+\nu p_\nu\,.
$$
The Poisson quadratic algebra satisfies the relations 
$$
\{K,R_1\}=2R_1,\qquad  
\{K,R_2\}=-2R_2,\qquad 
\{R_1,R_2\}=-8K^3-4aKH-16dK.
$$
These three extra constants are related via the identity 
$$
-R_1R_2 + K^4 + aHK^2 + 4dK^2 + H^2 = 0\,.
$$

This Hamiltonian admits a separation of variables in all the coordinate systems
that are possible.  We need only give the expressions in terms of
the fourth representatives. 
In the coordinate system associated with the fourth representative and for which 
$\gamma \neq 2$ the Hamiltonian can be written as 
\beann
H &=& 16XY \ \frac{X(A_+X-A_-)(A_-X-A_+)p^2_X-Y(A_+Y+A_-)(A_-Y+A_+)p^2_Y}
                  {(X-Y)(a-2-XY(a+2))A_+A_-} \\
 & & {} \ -\frac{4dA_+A_-\left(X^2Y+Y+XY^2+X\right)}{ (X-Y)(-a+2+XY(a+2))A_+A_-}\,,
\eeann
and for the case $\gamma =2$ this Hamiltonian has the form 
$$
H = \frac{\ds p_{\varphi'}^2+p_{\omega'}^2
            + d\left(\frac1{\sinh^22\omega'}+\frac1{\sin^22\varphi'}\right)}
         {\ds \frac{a+2}{\sinh^22\omega'}+\frac{a-2}{\sin^22\varphi'}}\,.
$$

The corresponding quadratic algebra relations are  
$$
[\hat K,\hat R_1]=2\hat R_1,\qquad 
[\hat K,\hat R_2]=-2\hat R_2,\qquad 
[\hat R_1,\hat R_2]=-8\hat K^3-4a\hat H\hat K-16d\hat K-4\hat K,
$$
subject to the operator identity 
$$
- \frac12\{\hat R_1,\hat R_2\} + \hat H^2 + a\hat H\hat K^2 + \hat K^4 + a\hat H
+ (5+4d)\hat K^2 + 4d = 0\,.
$$

 This completes the analysis of the superintegrable potentials associated with 
the four metrics of Darboux.

\section{Relationship to constant curvature superintegrable potentials}
\label{sec:CCM}

In sections \ref{sec:Darboux2}--\ref{sec:Darboux4}
we have found, by means of exhaustive calculation, all
superintegrable potentials in the Darboux spaces of revolution
having two or more quadratic integrals.
Once these are expressed in suitable coordinates, it is 
clear that each is simply a multiple of one of the superintegrable
potentials on the complex Euclidean plane or two-sphere, 
that have been enumerated in \cite{KKMP}, though that was by no means
evident in advance.

In each case we can start with a Hamiltonian of the form
\be
\label{H}
H = H_0 + \alpha V_0\,,
\ee
where $V_0$ is a function of the coordinates $x$ and $y$, and
$\alpha$ is a constant.
Dividing the Hamilton-Jacobi equation, $H=E$, 
throughout by $V_0$ and rearranging gives a new
Hamilton-Jacobi equation in which the roles of the energy $E$ and parameter
$\alpha$ have been exchanged.
\be
\label{Hdashed}
H' = \frac{H_0}{V_0} - \frac E{V_0} = -\alpha\,.
\ee
Clearly, the integrability and separability 
of one system guarantees that of the other.
It is this relationship between the harmonic oscillator potential 
written in Cartesian coordinates
and the Coulomb potential in parabolic coordinates that
has been discovered by many authors.  Transformations of this type
relating integrable systems were described in a more
general context by Hietarinta {\em et al} in \cite{HGDR} and called {\em coupling constant metamorphosis}. See also
\cite{BKM} where the {\em St\"ackel transform} and its close connection with variable separation was emphasized.

The preservation of integrability under such a transformation can be
demonstrated explicitly by noting that if $\{H_0,L_0\}=0$ and 
\be
\label{HforCCM}
H = H_0 + \alpha V_0 \quad \mbox{and} \quad L = L_0 + \alpha \ell_0
\ee
are in involution, i.e., $\{H,L\}=0$, then so are
\be
\label{CCM}
H' = \frac{H_0}{V_0} \quad \mbox{and} \quad L' = L_0 - \ell_0H'\,.
\ee

Any identities involving integrals
associated with (\ref{H}), give rise to corresponding identities
involving integrals associated with (\ref{Hdashed}) and are obtained
by the replacements 
\be
\label{CCMalphaH}
\alpha\rightarrow-H' \qquad \mbox{and} \qquad H\rightarrow0\,.
\ee

\subsection{Generating the Darboux spaces of revolution by coupling
constant metamorphosis}

Taking each of the degenerate potentials from \cite{KKMP}, that is, the
potentials with Hamiltonians having one first order and
two quadratic integrals and performing a coupling constant
metamorphosis we arrive at a Hamiltonian having
one first order $K$ and two quadratic constants, $X_1$ and $X_2$.  
These must be free
Hamiltonians either on one of the four Darboux spaces of revolution or
one of the constant curvature spaces,
$E_2(\mathbb C)$ or $S_2(\mathbb C)$.  After comparing the Hamiltonians
so generated, it can been seen that this approach generates all
of the Darboux spaces of revolution.

Knowing the Poisson algebra for each Hamiltonian
involved and how coupling constant metamorphosis
modifies this algebra, we can determine which
Hamiltonian has been generated, even if it
appears in unfamiliar coordinates.  Note that some transformations
reproduce the free Hamiltonian on $E_2(\mathbb C)$ or 
$S_{2,\mathbb C}$, and some Darboux spaces can be generated from two
distinct constant curvature potentials.

For each Hamiltonian we have four linearly independent constants
of the motion.  These, however, cannot be functionally
independent and there is always a polynomial identity in $K$, $X_1$,
$X_2$ and $H$ that is of fourth order in the momenta.  We can use this
identity to classify the possible Hamiltonians.
Up to freedoms in choosing $X_1$ and $X_2$, scalings of $K$ and
coupling constant metamorphosis, we find that there are 5 classes 
of identities that involve all of the constants.
The correspondences between these identities, degenerate superintegrable
potentials from \cite{KKMP} and the Darboux spaces of revolution 
are summarised in Table \ref{Hamtable}.  Note that because we
allow coupling constant metamorphosis, $H$ has the same status as 
parameters in the the potential and the coefficients $A$ and $B$
appearing in the representative identities may be functions of $H$.
The labels in bold
(e.g. {\bf E3}, {\bf S3},\ldots) refer to \cite{KKMP}.
Those Hamiltonians in Table \ref{Hamtable}
on the complex two-sphere, that is, {\bf S3}, {\bf S5} and {\bf S6},
are represented with three coordinates $s_1$, $s_2$ and $s_3$ 
constrained by 
$s_1^2+s_2^2+s_3^2=1$ and $J_1=s_2p_{s_3}-s_3p_{s_2}$,
$J_2=s_3p_{s_1}-s_1p_{s_3}$ and $J_3=s_1p_{s_2}-s_2p_{s_1}$.
The potentials {\bf E12}, {\bf E14}, {\bf E4} and {\bf E13} are functions
of $x-iy$ and hence division of $p_x^2+p_y^2$ by these potentials reproduces
the flat space Hamiltonian.

For example, starting from the algebraic identity for constants
associated with the Hamiltonian and integrals
\be
\label{E6}
H=p_x^2+p_y^2+\frac\alpha{x^2}+\alpha \qquad \mbox{({\bf E6})}\,,
\ee
$$
X_1 = (xp_y-yp_x)p_x - \frac{\alpha y}{x^2}\,, \qquad 
X_2 = (xp_y-yp_x)^2 + \frac{\alpha y^2}{x^2}\,, \qquad K=p_y\,,
$$
that is \cite{KKMP},
$$
X_1^2 + K^2X_2 - (H-\alpha)X_2 +\alpha K^2 = 0\,,
$$
we find that applying the transformation (\ref{CCM}) gives
$$
H' = \frac{p_x^2+p_y^2}{\frac1{x^2}+1}\,, \qquad 
X'_1 = (xp_y-yp_x)p_x - \frac{y}{x^2}\frac{p_x^2+p_y^2}{\frac1{x^2}+1}
     = \frac{y(p_y^2-x^2p_x^2)}{x^2+1} + xp_xp_y\,,
$$
$$ 
X'_2 = (xp_y-yp_x)^2 + \frac{y^2}{x^2}\frac{p_x^2+p_y^2}{\frac1{x^2}+1}
     = \frac{(x^2+x^4-y^2)p_y^2+x^2y^2p_x^2}{x^2+1}-2xyp_xp_y\,,\qquad K'=K\,,
$$
and using (\ref{CCMalphaH}),
$$
X_1^{\prime2} + K^{\prime2}X'_2 - H'X'_2 -H'K^{\prime2} = 0\,.
$$
Then
$$
X''_1=2X'_1\,,\qquad X''_2=-X'_2+H'\,, \qquad H''=H'\,,\qquad
K''=K'\,,
$$
gives
$$
X^{\prime\prime2}_1 - 4K^{\prime\prime2}X''_2 + 4H''X''_2 - 4H''^2 = 0\,,
$$
the identity (\ref{identity2}) associated with the Darboux space of type two
(\ref{H2}).

\begin{table}
\caption{Correspondences between constant curvature superintegrable potentials and
Hamiltonians for Darboux spaces of revolution.}
\label{Hamtable}
\begin{center}
\begin{tabular}{|rl|rl|c|}
\hline
\multicolumn{2}{|p{5cm}|}{\centering Degenerate superintegrable potential on 
                             $E_2(\mathbb C)$ or $S_2(\mathbb C)$}
  & \multicolumn{2}{p{5cm}|}{\centering Hamiltonian for Darboux space of revolution}
    & Representative identity \\ 
\hline \hline &&&& \\
{\bf E5}:  & $\ds 4x$ 
            & $D_1$: & $\ds \frac{p_u^2+p_v^2}{4u}$
              & $X_1^2 + A X_2 + K^4 + B = 0$ \\ &&&& \\
\hline &&&& \\
{\bf E6}:  & $\ds \frac1{x^2}+1$ 
            & $D_2$: & $\ds \frac{u^2(p_u^2+p_v^2)}{u^2+1}$
              & $X_1^2 + K^2X_2 + A X_2 + B = 0$ \\ &&&& \\ 
{\bf S5}:  & $\ds \frac1{(s_1-is_2)^2}-1$
            &&& \\ &&&& \\
\hline &&&& \\
{\bf E12}: & $\ds \frac{\alpha(x-iy)}{\sqrt{(x-iy)^2 + c^2}} + \beta$ 
            & $E_2(\mathbb C)$ &
               & $X_1^2 + K^2X_2 + A = 0$ \\ &&&& \\
{\bf E14}: & $\ds \frac\alpha{\sqrt{x-iy}} + \beta$ &&& \\ &&&& \\
\hline &&&& \\
{\bf E3}:  & $x^2+y^2+4$ 
            & $D_3$: & $\ds \frac{p_u^2+p_v^2}{4+u^2+v^2}$
                       & $X_1X_2 + A K^2 + B = 0$ \\ &&&& \\
{\bf E18}: & $\ds \frac2{\sqrt{x^2+y^2}} + 1$ 
             &&& \\ &&&& \\
\hline &&&& \\
{\bf S3}:  & $\ds \frac{a+2}{s_3^2} - a+2$ 
            & $D_4$: & $\ds \frac{p_u^2+p_v^2}{\frac{a+2}{u^2}+\frac{a-2}{v^2}}$
              & $X_1X_2 + K^4 + A K^2 + B = 0$ \\ &&&& \\
{\bf S6}:  & $\ds \frac{2is_3}{\sqrt{s_1^2+s_2^2}} + a$
             &&& \\ &&&& \\
\hline &&&& \\
{\bf E4}:  & $\ds \alpha(x-iy) + \beta$
            & $E_2(\mathbb C)$ &
              & $K^2X_1 + A X_2 + B = 0$ \\ &&&& \\
{\bf E13}: & $\ds \frac\alpha{\sqrt{x-iy}} + \beta$ &&& \\ &&&& \\
\hline
\end{tabular}
\end{center}
\end{table}

\subsection{Generating superintegrable potentials on Darboux spaces}

The $H_0$ in equation (\ref{HforCCM}) may itself contain potential terms
and if these are chosen so that $H$ is superintegrable, then so will be
$H'$.

For example, taking the superintegrable Hamiltonian on the complex 
two-sphere {\bf S1} \cite{KKMP},
$$
H = J_1^2 + J_2^2 + J_3^2 + \frac\alpha{(s_1-is_2)^2}
        + \frac{\beta s_3}{(s_1-is_2)^3}
        + \frac{\gamma(1-4s_3^2)}{(s_1-is_2)^4} + \delta
$$
and dividing though by $(s_1-is_2)^{-2}-1$ gives, after a change of
coordinates, the superintegrable potential {\bf [A]} in a Darboux
space of type 2.  The same Hamiltonian can be generated by 
dividing {\bf E2} throughout by $x^{-2}+1$.

Each potential in Table \ref{Hamtable} is compatible with the addition
of further terms while maintaining superintegrability, 
and in using the method demonstrated above, all 
superintegrable Hamiltonians found in Sections 
\ref{sec:Darboux2}--\ref{sec:Darboux4} can be generated.  The
correspondences are given below.

\subsubsection{Darboux spaces of type 1}
The potential {\bf E5}, $V_0 =4x$, appears in each of
\beann
\makebox[20mm]{\bf E2\hfill}    & : &
         \alpha(4x^2+y^2) + \beta x + \frac\gamma{x^2} + \delta \\
\makebox[20mm]{\bf E3$'$\hfill} & : &
         \alpha(x^2+y^2) + \beta x + \gamma y + \delta \\
\makebox[20mm]{\bf E9\hfill}    & : &
         \frac\alpha{\sqrt{x-iy}} + \beta x
           + \frac{\gamma(2x-iy)}{\sqrt{x-iy}} + \delta\,.
\eeann
The potential labelled {\bf E3$'$} is a translation of {\bf E3} . 
Adding these potentials to $H_0=p_x^2+p_y^2$ and dividing by $4x$ 
produces the two real non-degenerate potentials found in \cite{KKW} and an
additional complex one given in this paper. (The details of the quadratic
algebra and defining operators   for the Hamiltonian
derived from {\bf E9} can be computed   using
(\ref{Hdashed})).

\subsubsection{Darboux spaces of type 2}
The potentials {\bf E6} and {\bf S5} appear in each of following.
\beann
\makebox[10mm]{\bf E1\hfill}\makebox[10mm]{\bf [B]\hfill}     & : & 
         \alpha(x^2+y^2) + \frac\beta{x^2} + \frac\gamma{y^2} + \delta \\
\makebox[10mm]{\bf E2\hfill}\makebox[10mm]{\bf [A]\hfill}     & : & 
         \alpha(4x^2+y^2) + \beta x + \frac\gamma{y^2} + \delta \\
\makebox[10mm]{\bf E16\hfill}\makebox[10mm]{\bf [C]\hfill}    & : & 
         \frac1{\sqrt{x^2+y^2}}\left(\alpha + \frac\beta{x+\sqrt{x^2+y^2}}
           + \frac\gamma{x-\sqrt{x^2+y^2}}\right) + \delta \\
\makebox[10mm]{\bf S1\hfill}\makebox[10mm]{\bf [A]\hfill}     & : & 
         \frac\alpha{(s_1-is_2)^2} + \frac{\beta s_3}{(s_1-is_2)^3}
           + \frac{\gamma(1-4z^2)}{(x-iy)^4} + \delta \\
\makebox[10mm]{\bf S2\hfill}\makebox[10mm]{\bf [B]\hfill}     & : & 
         \frac\alpha{s_3^2} + \frac\beta{(s_1-is_2)^2}
           + \frac{\gamma(s_1+is_2)}{(s_1-is_2)^3} + \delta \\
\makebox[10mm]{\bf S4\hfill}\makebox[10mm]{\bf [C]\hfill}     & : &
         \frac\alpha{(s_1-is_2)^2} + \frac{\beta s_3}{\sqrt{s_1^2+s_2^2}}
           + \frac\gamma{(s_1-is_2)\sqrt{s_1^2+s_2^2}} + \delta
\eeann
The superintegrable system generated after dividing by $x^{-2}+1$ or
$(s_1-is_2)^{-2}-1$ as appropriate is indicated by label the {\bf [A]},
{\bf [B]} or {\bf [C]}.
The apparent over abundance of superintegrable potentials 
generated in this way for $D_2$ is
resolved by noting that the same potential can appear in more than
one coordinate system.

\subsubsection{Darboux spaces of type 3}
The potentials {\bf E3} and {\bf E18} appear in each of
\beann
\makebox[10mm]{\bf E1\hfill}\makebox[10mm]{\bf [B]\hfill}     & : &
         \alpha(x^2+y^2) + \frac\beta{x^2} + \frac\gamma{y^2} + \delta \\
\makebox[10mm]{\bf E3$'$\hfill}\makebox[10mm]{\bf [A]\hfill}  & : &
         \alpha(x^2+y^2) + \beta x + \gamma y + \delta \\ 
\makebox[10mm]{\bf E7\hfill}\makebox[10mm]{\bf [D]\hfill}     & : &
         \frac{\alpha(x-iy)}{\sqrt{(x-iy)^2-c^2}}
           + \frac{\beta(x+iy)}{\sqrt{(x-iy)^2-c^2}\left((x-iy)+\sqrt{(x-iy)^2-c^2}\right)^2} \\
 & & \qquad 
           + \gamma(x^2+y^2) + \delta \\ 
\makebox[10mm]{\bf E8\hfill}\makebox[10mm]{\bf [C]\hfill}     & : &
         \frac{\alpha(x+iy)}{(x-iy)^3} + \frac\beta{(x-iy)^2}
           + \gamma(x^2+y^2) + \delta \\
\makebox[10mm]{\bf E16\hfill}\makebox[10mm]{\bf [B]\hfill}     & : &
         \frac1{\sqrt{x^2+y^2}}\left(\alpha + \frac\beta{x+\sqrt{x^2+y^2}}
           + \frac\gamma{x-\sqrt{x^2+y^2}}\right) + \delta \\
\makebox[10mm]{\bf E17\hfill}\makebox[10mm]{\bf [C]\hfill}     & : &
         \frac\alpha{\sqrt{x^2+y^2}} + \frac\beta{(x+iy)^2}
           + \frac\gamma{(x+iy)\sqrt{x^2+y^2}} + \delta \\
\makebox[10mm]{\bf E19\hfill}\makebox[10mm]{\bf [D]\hfill}     & : &
         \frac{\alpha(x-iy)}{\sqrt{(x-iy)^2-4}}
           + \frac\beta{\sqrt{(x+iy)(x-iy+2)}} \\
 & & \qquad 
           + \frac\gamma{\sqrt{(x+iy)(x-iy-2)}} + \delta \\
\makebox[10mm]{\bf E20\hfill}\makebox[10mm]{\bf [A]\hfill}     & : &
         \frac1{\sqrt{x^2+y^2}}\left(\alpha 
           + \beta\sqrt{x+\sqrt{x^2+y^2}}
           + \gamma\sqrt{x-\sqrt{x^2+y^2}}\right) + \delta \\
\eeann
As before, once the possibility of changes of coordinates 
in taken into account,
the above list produces only those superintegrable potentials 
found in section \ref{subsec:superint3}.

\subsubsection{Darboux spaces of type 4}
The potentials {\bf S3} and {\bf S6} appear in each of
\beann
\makebox[10mm]{\bf S2\hfill}\makebox[10mm]{\bf [A]\hfill}     & : &
         \frac\alpha{s_3^2} + \frac\beta{(s_1-is_2)^2}
           + \frac{\gamma(s_1+is_2)}{(s_1-is_2)^3} + \delta \\
\makebox[10mm]{\bf S4\hfill}\makebox[10mm]{\bf [A]\hfill}     & : &
         \frac\alpha{(s_1-is_2)^2} + \frac{\beta s_3}{\sqrt{s_1^2+s_2^2}}
           + \frac\gamma{(s_1-is_2)\sqrt{s_1^2+s_2^2}} + \delta \\
\makebox[10mm]{\bf S7\hfill}\makebox[10mm]{\bf [B,C]\hfill}     & : &
         \frac{\alpha s_1}{\sqrt{s_2^2+s_3^2}} + \frac{\beta s_2}{s_3^2\sqrt{s_2^2+s_3^2}}
           + \frac\gamma{s_3^2} + \delta \\
\makebox[10mm]{\bf S8\hfill}\makebox[10mm]{\bf [C]\hfill}     & : &
         \frac{\alpha s_1}{\sqrt{s_2^2+s_3^2}} + \frac{\beta(s_1+is_2-s_3)}{\sqrt{(s_1+is_2)(s_3-is_2)}}
           + \frac{\gamma(s_1+is_2+s_3)}{\sqrt{(s_1+is_2)(s_3+is_2)}} + \delta \\
\makebox[10mm]{\bf S9\hfill}\makebox[10mm]{\bf [B]\hfill}     & : &
         \frac\alpha{s_1^2} + \frac\beta{s_2^2} + \frac\gamma{s_3^2} + \delta \\
\eeann
As before, once the possibility of changes of coordinates 
in taken into account,
the above list produces only those superintegrable potentials 
found in section \ref{subsec:superint4}.

\section{Conclusion}

In this paper we have discussed in some detail three of the four Darboux spaces
of revolution that have at least two integrals of classical motion quadratic in
the momenta in addition to the Hamiltonian.  In each case we have also presented 
an exhaustive list of potentials for each of these spaces which when added to 
the Hamiltonians of these spaces preserve this property i.e.\ that there are 
still two extra integrals of the classical motion. These are the superintegrable
systems associated with the systems of Darboux. The property of extra integrals 
also extends easily to the case of the corresponding quantum systems. For each 
of these systems we have calculated the corresponding quadratic algebra 
relations and shown that in each case the Hamiltonians that we obtain arise 
from constant curvature systems via a coupling constant transformation. We have 
also discussed the solutions of the corresponding classical and quantum 
problems in each of the inequivalent coordinate systems and have also given 
some of the embeddings of these spaces in three dimensions. In the last section 
we have shown how the free Hamiltonians of Darboux are related to the 
superintegrable Hamiltonians on spaces of constant curvature via coupling 
constant transformations. We also list how the corresponding superintegrable 
systems of spaces of constant curvature are related in this way to the 
superintegrable systems that we have found. This classification is comprehensive
and complete.

Let us very briefly review the current status of superintegrability
in two-dimensional spaces. Most of the published work 
\cite{FMSUW,WSUF,KMJP2,KMJP3,KKMP} concerns 
quadratic superintegrability for classical, or quantum Hamiltonians of the 
form kinetic energy plus a scalar potential. Once a specific space is 
chosen, superintegrable systems in the space can be classified under 
the action of the corresponding isometry group. Systems in the same class
are not only mathematically equivalent, but also have the same physical 
properties. In classical mechanics they will have the same trajectories
and the trajectories will be periodic, if they are bounded. Similarly,
in quantum mechanics superintegrable systems in the same class will have 
the same energy levels and eigenspaces.

Quadratically superintegrable systems exist in spaces of constant 
curvature and also in Darboux spaces. A Darboux space is defined by the 
fact that it allows one Killing vector and two (irreducible) Killing 
tensors. This article completes the task of classifying all quadratically
superintegrable systems in all of the above spaces. 

The results are quite rich. Indeed, in the real Euclidean space $E_2$, we
have four $E(2)$ classes of superintegrable systems \cite{FMSUW,WSUF}. They are
physically quite diverse. One is an isotropic harmonic oscillator with
additional terms, called {\bf E1} above in Section \ref{sec:CCM}. 
A second is an anisotropic
harmonic oscillator with additional terms (called {\bf E2} above). The third
and fourth are Kepler (or Coulomb) systems with two different types of
additional terms, respectively. In complex Euclidean space $E_2(\mathbb C)$, or
correspondingly in the pseudo-Euclidean space $E(1,1)$, one obtains 6 more
classes \cite{KMJP2}.

Two classes of superintegrable systems exist on the real sphere $S_2$,
four more on the complex sphere $S_2(\mathbb C)$ \cite{KMJP3}. 
On the real Darboux spaces 
$D_1,\ldots,D_4$ we have obtained 3, 4, 4, and 4 classes of systems, 
respectively. One more for the complex space $D_3(\mathbb C)$.

From the mathematical point of view the situation is much more unified.
As was stressed above, superintegrable systems that may correspond to 
quite different physical situations may be related by coupling constant 
metamorphosis. Once we allow this type of equivalence, many fewer 
equivalence classes exist. For instance, in real Euclidean space we only 
have two classes, because the Kepler potentials with additional terms
are equivalent to isotropic harmonic oscillators (in one case with the 
additional terms). All superintegrable systems in Darboux spaces are 
related by coupling constant metamorphosis to systems in spaces of 
constant curvature. For $D_1$, $D_2$ and $D_3$ this is always flat space, 
complex or real. Two of the systems in $D_4$ are related to systems in
real Euclidean space. The other two are related to systems on a complex 
sphere. The relation is of course not unique and depends on the choice of 
coordinates (see Section \ref{sec:CCM}).

A typical feature of quadratic superintegrability for scalar potentials
is that quantum and classical superintegrable potentials coincide. They 
allow separation of variables in at least two coordinate systems in the 
Schr\"odinger and Hamilton-Jacobi equation, respectively.

Superintegrability involving third order integrals of motion has 
also been considered \cite{GW,G}. There the situation is quite different.
Multiseparability is lost. More interestingly, quantum superintegrable 
systems exist (in real Euclidean space) that have no classical analog
(in the classical limit they reduce to free motion).

\section*{Acknowledgements}
The research of P.~W.\ was partly supported by research grants from NSERC
of Canada and FQRNT du Quebec.  J.~K.\ began this work while supported by
the  contract``special functions, superintegrability and
separation of variables" of the New Zealand Marsden Fund.

\end{document}